# Spectral investigation of Ceres analogue mixtures: in-depth analysis of crater central peak material (ccp) on Ceres

Galiano A. (1), Dirri F. (1), Palomba E. (1,2), Longobardo A. (1), Schmitt B. (3), Beck P. (3)

(1) INAF-IAPS, Rome, Italy (anna.galiano@inaf.it), (2) SSDC-ASI, Rome, Italy, (3) Université Grenoble Alpes, CNRS, IPAG, F-38000 Grenoble, France.

## Abstract

The dwarf planet Ceres is an airless body composed of Mg-phyllosilicates, $NH_4$-phyllosilicates, and Mg/Ca-carbonates, in addition to a dark component. The subsurface of Ceres, investigated by the material composing the peak of complex craters (ccp, crater central peak material; *Galiano et al., 2019*), reveals a composition similar to the surface, with an increasing abundance of phyllosilicates in the interior. A moderate trend between age of craters' formation and spectral slope of ccps suggests that younger ccps show a negative/blue slope and older ccps are characterized by positive/red slope. Different hypothesis have been proposed to explain such spectral trend, including variation in composition and/or grain size. In order to investigate the causes of different spectral slope in ccps, different grain-sizes of Ceres analogue mixtures were produced, heated to remove absorption of atmospheric water, and spectrally analyzed. First, the end-members which compose the Ceres surface (using the antigorite as Mg-phyllosilicate, the $NH_4$-montmorillonite as $NH_4$-phyllosilicate, the dolomite as carbonate and the graphite as dark component), were mixed, obtaining mixtures with different relative abundance, and identifying the mixture with the reflectance spectrum most similar to the average Ceres spectrum. The mixtures were obtained with grain size of 0-25 μm, 25-50 μm and 50-100 μm, were heated and spectrally analyzed, both with an acquisition temperature of 300 K (room temperature) and 200 K (typical for surface Ceres temperature during VIR observations).

The 3.1 and 3.4 μm bands in $NH_4$-montmorillonite spectra are weaker at increasing temperatures, related, respectively, to the devolatilization of OH group and the enlargement of atomic distances. The 2.7 and 3.4 μm band depth are deeper in coarser samples of $NH_4$-montmorillonite, related to the volume scattering which occurs among grains. An opposite trend is instead observed for the 3.1 μm band in $NH_4$-montmorillonite sample and for the 2.7 μm of Antigorite sample, explained with bands' saturation. From the analysis of different Ceres analogue mixtures, the 3.1 μm band center shifts toward longer wavelength at increasing abundances of $NH_4$-montmorillonite.

After the application of heating process, the 3.4 and 4.0 μm bands show deeper intensity, as well as the 2.7 and the 3.1 μm band, likely due to the loss of absorbed atmospheric water: the latter process is also responsible of the narrowing of 3.1 µm band and the shift of band center toward longer wavelength, i.e. at 3.06 µm, coincident with mean Ceres spectrum. Furthermore, a darkening and reddening of spectrum is observed after the heating process, as consequence of the devolatilization of OH group in phyllosilicates and a more dominant effect of opaque phase.

The analysis of the most similar Ceres analogue mixture, reproduced at different grain size and underwent to heating process, reveal a weakening of 2.7, 3.1, 3.4 and 4.0 absorption bands in coarser samples, likely related to large size of dark grains which reduce the spectral contrast. Furthermore, a new result emerges between spectral slope and grain size: spectra show a more positive slope in coarser mixtures, probably influenced by the trend of larger size of dark component.

The most similar Ceres analogue mixture is composed of dolomite (18%), graphite (27%), antigorite (32%) and $NH_4$-montmorillonite (29%) and the results of this work suggest that this mixture is more similar to the Ceres youngest region (such as Haulani) than to the Ceres average, in particular for the negative slope of spectrum.

In order to obtain a mixture able to reproduce the mean Ceres, some suggestions are provided. Small variation in the composition and grain size of end-members need to be considered, in addition to the occurrence of a dark component dispersed in fine size. Furthermore, the positive spectral slope that characterizes the mean Ceres spectrum can be obtained by the application of some processes simulating the space weathering on Ceres (as micro-meteoritic impacts and solar wind irradiation), i.e. laser and ion irradiation.

As conclusion, youngest ccps on Ceres (as Haulani) are probably composed by fresher and weakly processed mixture with fine dark material intimately dispersed: as a result, the reflectance spectra of youngest material show a negative slope in the 1.2-1.9 µm range. The redder slope observed in the older ccps is probably the consequence of the space weathering effects on fresher material.

# 1. Introduction

The dwarf planet Ceres has been visited by the Dawn spacecraft from April 2015 to October 2018. During three years of observations, images, hyperspectral data and elemental composition of Ceres surface were acquired with increasing resolutions by using the Framing Camera (FC; *Sierks et al., 2011*), the Visible and Infrared Mapping Spectrometer (VIR; *De Sanctis et al., 2011*), the Gamma Ray and Neutron Detector (GRaND; *Prettyman et al., 2011*). VIR data revealed a global distribution of phyllosilicates over the surface *(Ammannito et al., 2016)*, in particular Mg-phyllosilicates and $NH_4$-phyllosilicates, mixed with Mg/Ca-carbonates and a dark featureless component, i.e., magnetite and/or graphite *(De Sanctis et al., 2015)*. In particular, reflectance spectra of Ceres surface show absorption bands at about 2.7 μm (due to Mg-phyllosilicates), at 3.1 μm ($NH_4$-phyllosilicates), at 3.4 μm and 3.9 μm (mainly due to carbonate). The low visual albedo value of about 0.033 is indicative of the occurrence of a dark component widespread in the Ceres regolith *(De Sanctis et al., 2015)*.

Bright and localized areas, termed as *Bright Spots* (BS) have been observed on the Ceres surface, characterized by a higher reflectance than the Ceres surface and by spectral variations, indicative of a slightly different composition *(Palomba et al., 2018)*. The brightest BS lies on the dome of the 92-km in diameter Occator crater, composed of Al-phylloslilicates, Na-carbonate, ammoniated salts and dark material *(De Sanctis et al., 2016)*.

The subsurface of Ceres has been investigated by analyzing central peak of complex craters, a geologic unit termed as *crater central peak material* (ccp) (*Galiano et al., 2019*): ccp is the subsurface and fresher material emerged as a consequence of impact craterinng. From the spectral analysis of 32 ccps, more detailed information about the composition of Ceres subsurface has been constrained, at depths included between 110 m and 22 km. At increasing depth of excavation, the abundance of Mg-phyllosilicates and ammoniated phyllosilicates increases (*Galiano et al., 2019*), supporting the existence of a silicate-rich mantle at depth of about 41 km *(Ermakov et al., 2017)*. Furthermore, a moderate trend has been observed between the spectral slope of ccps (estimated between 1.2 and 1.9 μm) and the age of craters' formation: older ccps are characterized by a positive spectral slope, whereas the youngest crater, Haulani, formed about 2.7 Myr ago *(Schmedemann et al., 2016)* shows a spectral slope of -0.067 *(Galiano et al., 2019)*.

An explanation of the trend observed between spectral slope and age of formation could be the space weathering processes *(Galiano et al., 2019)* that, in an airless body as Ceres, are due to solar wind irradiation *(Gillis-Davis, 2016)*, micro-meteoritic hypervelocity impacts *(Pieters and Noble, 2016)*, mixing of materials by impact and thermal fatigue *(Delbo et al., 2014)*.

Furthermore, laboratory measurements demonstrated that coarse samples show bluer (negative) slope than fine grains *(Cloutis et al., 2011)*: youngest ccps could be composed by coarse grains, which could be fragmented over time by micro-meteoritic impacts and/or by thermal fatigue. Anyway, also the chemical composition could produce different spectral slopes and phenomena linked to impact which generated complex craters could affect the material composing the peak, resulting in variation of spectral slope.

In order to better understand the trend assumed by spectral slope in ccps, different grain-sized Ceres analogue mixtures were produced and spectrally analysed. The spectra of mixtures were acquired both at room temperature (290-300 K) and at cold temperatures (i.e. 200 K), since the Ceres surface daily temperature, estimated at the equator and measured at 2.86 AU from the Sun, ranges from 180 K to 240 K *(De Sanctis et al., 2015)*.

In the Chapter 2, the experimental procedure is described, and the Chapter 3 is dedicated to the spectral analysis performed on the samples. The Chapter 4 includes the spectral analysis of end-members used to produce Ceres analogue mixtures and in Chapter 5 the Ceres analogue mixtures are spectrally investigated. In the Chapter 6, the results are discussed, and the conclusions are given in the Chapter 7.

## 2. Experimental procedure and instrumentation

The mixtures were prepared and analyzed at the laboratory "Cold Surface Spectroscopy Facility" at the Institute de Planétologie et Astrophysique de Grenoble (IPAG). The mixtures were produced with the four end-members suggested to compose the Ceres surface *(De Sanctis et al., 2015)*, i.e. Mg-/Ca-carbonate (dolomite), Mg-phyllosilicate (antigorite), $NH_4$-phyllosilicate ($NH_4$-montmorillonite), dark component (graphite). Each end-member was first ground in a pestle and mortar, obtaining grain size greater than 100 μm. Then, by using an automatic sieving system, three different grain sizes were obtained for each end-member: 0-25 μm, 25-50 μm, 50-100 μm.

In order to detect the most similar Ceres analogue mixture, the end-members were mixed by varying their abundance, and 6 mixtures were obtained (the details about mineral abundances for each mixture is in *Table 2*): Mixture #1 (grain size <25 μm), Mixture #2 (grain size 50-100 μm), Mixture #3a (grain size 50-100 μm), Mixture #3b (grain size 50-100 μm), Mixture #4 (grain size 50-100 μm). To evaporate the absorbed atmospheric water, the Mixture #4 was heated in oven with a temperature range included between 390 K and 430 K for about two hours, and the Mixture #4_h (grain size 50-100 μm) was obtained.

The bidirectional reflectance spectra of mixtures were acquired in the VIS-NIR spectral range between 0.4 and 4.2 μm, at room temperature (290 K) in vacuum chamber and with an incidence angle *i* of 0° and an emission angle *e* of 30°. The mixture most spectrally similar to Ceres was found to be Mixture #4_h, which was reproduced at grain size 0-25 μm and 25-50 μm and heated in oven for about two hours at 370-390 K. Therefore, the three heated samples of the selected mixture (grain size 0-25 μm, 25-50 μm and 50-100 μm) were analyzed by acquiring their bidirectional spectra in vacuum chamber at cold and room temperatures (from 180 K to 300 K) to better simulate the Ceres environment, with the viewing angles previously used ($i$=0° and $e$ =30°).

The spectra of the four end-members were also acquired at different grain sizes in the same spectral range and temperature, in order to reveal the trend assumed by spectral parameters of pure minerals at varying grain size and temperatures.

The facility hosts two different instrument: the SHADOWS Micro Spectro-Gonio Radiometer (Spectro-photometer with cHanging Angles for Detection Of Weak Signals) *(Potin et al., 2018)* and the SHINE Spectro-Gonio Radiometer (SpectropHotometer with variable INcidence and Emergence) *(Bonnefoy, 2001; Brissaud et al., 2004)*. The bidirectional reflectance spectrum of Mixture #1 was acquired with the SHADOWS Micro Spectro-Gonio Radiometer, optimized for particularly dark samples. The spectra of all the other mixtures and end-members were acquired by using the SHINE Spectro-Gonio Radiometer. The spectral range of SHADOWS is between 0.4 and 4.7 μm, and the spectral range of SHINE is between 0.35 and 4.5 μm. The reflectance spectra acquired in this work range from 0.4 to 4.2 μm with the following spectral resolutions: 6 nm (from 0.4 to 0.7 μm); 12 nm (from 0.75 to 1.5 μm); 24 nm (from 1.5 to 3.0 μm); 48 nm (from 3.0 to 4.2 μm).

The instruments are located in a dark cold room that can be cooled down to 250 K. The SHINE instrument, in particular, was used to acquire spectra of mixtures at cold temperature (to simulate the Ceres environment), since it can be cooled down to 70 K by using CarboN-IR environmental cell *(Grisolle, 2013; Beck et al. 2015)*.

## 3. Tools

As the radiation interacts with an optically thick sample, it can be either reflected or absorbed *(Wendlandt and Hecht, 1966)*. The reflected radiation can be characterized by a surface reflection (surface scattering) and a diffuse reflection (volume scattering), in which some of the incident light penetrates the sample and it is partially absorbed before being scattered back to the surface. For transparent or weakly absorbing materials, spectra are dominated by diffuse reflection: in finer samples the reflection increases, reducing the depth of penetration of incident light and weaker absorptions are obtained; in coarse-grained samples, photons have low probability to be scattered out of the sample, therefore the depth of penetration of light into the sample increases and deeper absorption bands occur *(Salisbury et al., 1993)*. Opaque minerals are instead dominated by surface scattering *(Cloutis et al., 2008)* since most of the radiation is absorbed by the particulate surface *(Salisbury et al., 1987)*: as consequence, a featureless reflectance spectrum is indicative of opaque phase. The amount of light absorbed in a surface is therefore dependent of the mean path length traversed through the material, which is in turn affected by the grain size and absorption coefficient *(Clark and Rousch, 1984)*. Even the temperature can alter the absorption of light by the sample *(Salisbury et al., 1993)*.

The NIR range included between 2 and 8 μm is characterized by a low absorption coefficient, resulting in the volume scattering of radiation through grains: photons penetrating the volume of grains can survive, passing through them and interacting with neighbor grains. Usually, at decreasing grain size, the elevated volume scattering produce weaker absorption bands. Sometimes, an opposite trend is observed in hydrated minerals, carbonates and silicates (for the O-H stretch, the C-O and Si-O overtone bands): the bands in finest samples are deeper than in coarser samples, since the band center is saturated, while reflectance in the and wings is controlled by the reduction of mean optical path length through the grains *(Salisbury et al., 1993)*. Furthermore, laboratory measurements demonstrated that coarse samples show bluer (negative) slope than fine grains *(Cloutis et al., 2011)*. Temperature variations can affect the molecular vibrations, producing a shift of band's position and a reduction in band intensity at rising temperatures *(Freund et al., 1974)*. At more elevated temperatures, the mean distances between atoms in a solid increase, reducing the dipole moment of molecule: as consequence, generally, the infrared bands are weaker at increasing temperatures *(Fruwert et al., 1966; Newszmelyi and Imre, 1968)* and band width becomes broader *(Freund et al., 1974, Salisbury et al., 1993)*. However, some exceptions occur: the OH-stretching band in kaolinite spectra becomes narrower at increasing temperatures, likely for an enlargement of interatomic distances related to thermal expansion of lattice, or to a weakening of some H-bond interaction between the hydroxyl groups *(Freund et al., 1974)*. Note also that some combination modes can also decrease in intensity with decreasing temperature (Beck et al., 2015).

The spectral trend shown with varying grain sizes and temperatures is investigated by defining band descriptors. The latter are usually calculated on the isolated bands, i.e. after fitting the spectral continuum by linear straight lines and removed *(Galiano et al., 2018)*. In this work we used the following band descriptors: band center (BC), band depth (BD), full width half maximum (FWHM). The BC is the wavelength corresponding to the minimum reflectance value in the isolated band. The BD is estimated as $1-R_{bc}/R_c$, where $R_{bc}$ is the reflectance at the band center and $R_c$ is the corresponding continuum reflectance *(Clark and Roush, 1984)*. The FWHM is the width at half the height between the band center and the continuum. The reflectance value at 2.1 μm ($R_{2.1}$) and the spectral slope estimated between 1.2 and 1.9 μm are other spectral parameters used in the analysis. The spectral slope is defined as follows: $[(R_{1.9} - R_{1.2})/ R_{1.9}]/(1.9-1.2)$.

To define relationships between spectral parameters, the Pearson correlation coefficient was considered (PCC). PCC estimates a linear correlation between two parameters by considering the covariance of variables ($cov(X,Y)$) and their standard deviations ($\sigma_X$ and $\sigma_Y$). Given N observations of variables **X** and **Y** and being $\bar{x}$ and $\bar{y}$ their mean values and $x_i$ and $y_i$ the single observations, the $cov(X,Y)$, $\sigma_X$ and $\sigma_Y$ are:

$$cov(X,Y) = \frac{1}{N}\sum_{i=1}^{N}(x_i - \bar{x})(y_i - \bar{y})$$

$$\sigma_X = \sqrt{\frac{\sum_{i=1}^{N}(x_i - \bar{x})}{N}} \qquad \sigma_Y = \sqrt{\frac{\sum_{i=1}^{N}(y_i - \bar{y})}{N}}$$

The definition of PCC is, consequently:

$$PCC = \frac{cov(X,Y)}{\sigma_X \sigma_Y} = \frac{\sum_{i=1}^{N}(x_i - \bar{x})(y_i - \bar{y})}{\sqrt{\sum_{i=1}^{N}(x_i - \bar{x})\sum_{i=1}^{N}(y_i - \bar{y})}}$$

PCC's values range from -1 (suggesting an anti-correlation) to 1 (indicative of a linear correlation): Pearson values included between 0 and 0.3 (or between -0.3 and 0) suggest a weak correlation (anti-correlation); values between 0.3 and 0.7 (or between -0.7 and -0.3) are indicative of a moderate correlation (anti-correlation); a strong correlation (or anti-correlation) is obtained for values between 0.7 and 1 (or between -1 and -0.7). A Pearson value close to 0 indicates an absence of correlation between the two considered spectral parameters.

## 4. Spectral analysis of end-members

The bidirectional reflectance spectra of end-members were acquired for different grain sizes and temperatures (*Table 1*). In particular, the bidirectional spectra of $NH_4$-montmorillonite (grain size 0-25 μm) and antigorite (50-100 μm) were acquired at cold and room temperatures, included between 180 and 290 K, while the dolomite (grain size 0-25 μm) was acquired at a room temperature of 300 K and the spectra of $NH_4$-montmorillonite (25-50 μm), antigorite (25-50 μm) and dolomite (50-100 μm) were acquired at 200 K. The reflectance spectrum of graphite, selected as dark component for this work, is flat and featureless. A representative spectrum of graphite was acquired at a room temperature of 290 K for a sample characterized by grain size greater than 100 μm. Representative spectra of the four end-members are shown in *Figure 1*, i.e. antigorite sample of 25-50 μm (spectrum acquired at 200 K), dolomite sample of 50-100 μm (spectrum acquired at 200 K), $NH_4$-montmorillonite sample of 25-50 μm (spectrum acquired at 200 K) and graphite sample with grain size greater than 100 μm (spectrum acquired at 290 K).

| Sample | Grain size (μm) | Temperature (K) |
| --- | --- | --- |

| End-member | Grain size | Temperature |
|---|---|---|
| NH$_4$-montmorillonite | 0-25 | 180 |
| NH$_4$-montmorillonite | 0-25 | 190 |
| NH$_4$-montmorillonite | 0-25 | 200 |
| NH$_4$-montmorillonite | 0-25 | 210 |
| NH$_4$-montmorillonite | 0-25 | 220 |
| NH$_4$-montmorillonite | 0-25 | 230 |
| NH$_4$-montmorillonite | 0-25 | 240 |
| NH$_4$-montmorillonite | 0-25 | 250 |
| NH$_4$-montmorillonite | 0-25 | 260 |
| NH$_4$-montmorillonite | 0-25 | 290 |
| NH$_4$-montmorillonite | 25-50 | 200 |
| Antigorite | 25-50 | 200 |
| Antigorite | 50-100 | 180 |
| Antigorite | 50-100 | 190 |
| Antigorite | 50-100 | 200 |
| Antigorite | 50-100 | 210 |
| Antigorite | 50-100 | 220 |
| Antigorite | 50-100 | 230 |
| Antigorite | 50-100 | 240 |
| Antigorite | 50-100 | 250 |
| Antigorite | 50-100 | 260 |
| Antigorite | 50-100 | 290 |
| Dolomite | 0-25 | 300 |
| Dolomite | 50-100 | 200 |
| Graphite | >100 | 290 |

*Table 1:* Each column describes the analyzed end-member, the grain size of the sample and the temperature at which the reflectance spectrum of the sample was acquired.

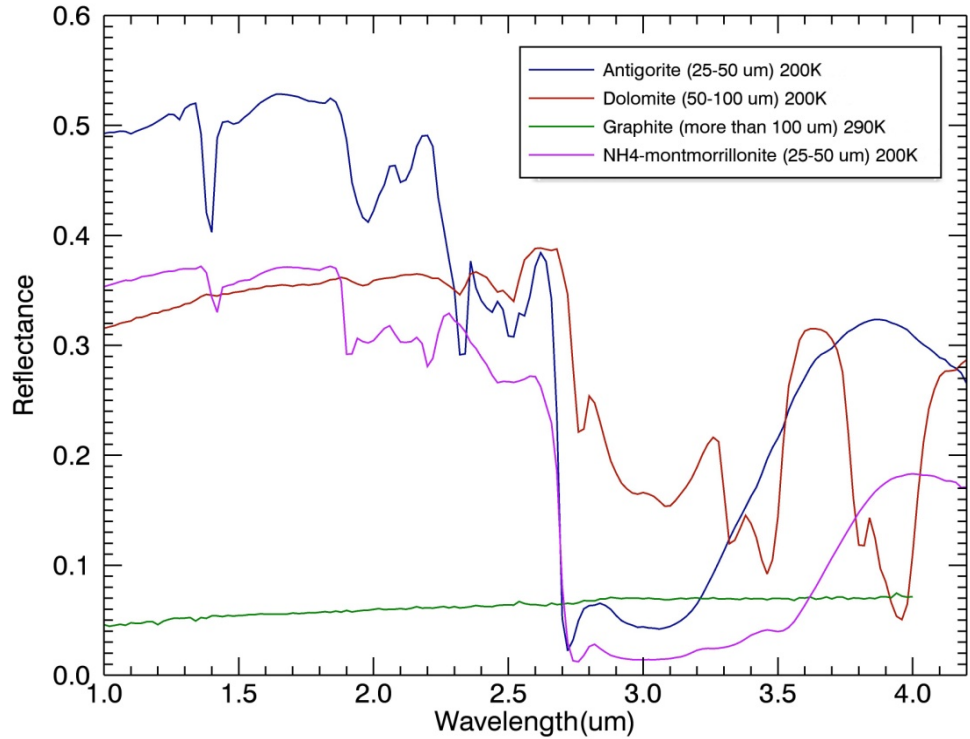

***Figure 1:*** Representative spectra of the four end-members suggested to compose the Ceres surface, i.e. antigorite (Mg-phyllosilicate), NH$_4$-montmorillonite (NH$_4$-phyllosilicate), dolomite (Mg-/Ca-carbonate), graphite (dark component).

### *4.1 NH$_4$-montmorillonite*

The NH$_4$-montmorillonite sample used in this work is a clay mineral which incorporated ammonium through laboratory process *(Busenberg and Clemency, 1973; Borden and Giese, 2001)*.

The NH$_4$-phyllosilicates are characterized by NH$_4$ incorporated in the phyllosilicate mineral lattice, present in an interlayer site to provide charge balance or, even, the NH$_4$ could be absorbed in the mineral surface *(Ehlmann et al., 2018)*. Free $NH_4^+$ shows the fundamental vibrational modes at about 3.29 μm (the symmetric stretch $v_1$), at about 5.95 μm (the in-plane bend $v_2$), at about 3.18 μm (the asymmetric stretch $v_3$) and at about 7.14 μm (the out-of-plane bend $v_4$), but only the $v_3$ and $v_4$ are IR active *(Harlov et al., 2001a)*. In particular, the absorption band due to $v_3$ fundamental appears in the 2.99-3.23 μm range and in the case of NH$_4$ incorporated in interlayers of phyllosilicates, the absorption appears in the 3.02-3.08 μm range *(Ehlmann et al., 2018)*, whereas secondary absorption bands can be observed in the 3.23-3.33 μm range (due to $v_2 + v_4$ combination) and at about 3.57 μm (attributable to $2v_4$) *(Harlov et al., 2001a,b,c; Boer et al., 2007)*. Anyway, the hydroxyl group (OH) composing the phyllosilicates is responsible of the absorption bands located at about 1.4 μm (overtone

O-H stretch $2\nu_3$) *(Clark, 1999)*, at about 2.22 μm (combination tone of Metal-O-H bend with O-H stretch) *(Gaffey et al., 1993)* and at about 2.7 μm ($\nu_1$ symmetric stretch of O-H group) *(Farmer, 1968)*. The $NH_4$-montmorillonite analyzed in this work shows absorption bands at about 1.4, 2.2, 2.7, 3.1 and 3.4 μm (**Figure 2**), in addition to a 1.9 μm spectral feature. The 1.9 μm absorption band is related, in phyllosilicates, to the combination of H-O-H $\nu_2$ bend and O-H $\nu_3$ asymmetric stretch *(Clark, 1999)*. Consequently, it is likely that atmospheric water was absorbed in the $NH_4$-montmorillonite sample and which provided a contribution in the 3.1 μm band *(Beck et al., 2010)*.

In this analysis, and in order to better discriminate the spectral behavior in Ceres mixture, the absorption band at about 3.1 μm (due to $\nu_3$ mode and probably to absorbed atmospheric water) and the overlapped bands between 3.23 and 3.57 μm (indicated as the 3.4 μm band) were taken into account for the $NH_4$-montmorillonite samples. Furthermore, also the 2.7 μm band ($\nu_1$ symmetric stretch of O-H group) was considered, since it is a spectral feature distinctive of phyllosilicate minerals. The 2.7, 3.1 and 3.4 μm bands were isolated by removing the spectral continuum. The spectral continuum was fitted by straight lines whose endpoints were estimated as local maxima in the following spectral ranges: 2.5-2.7 μm and 2.8-2.9 μm (for the 2.7 μm band); 2.8-2.9 μm and 3.15-3.3 μm (for the 3.1 μm band); 3.15-3.3 μm and 3.8-4.1 μm (for the 3.4 μm band).

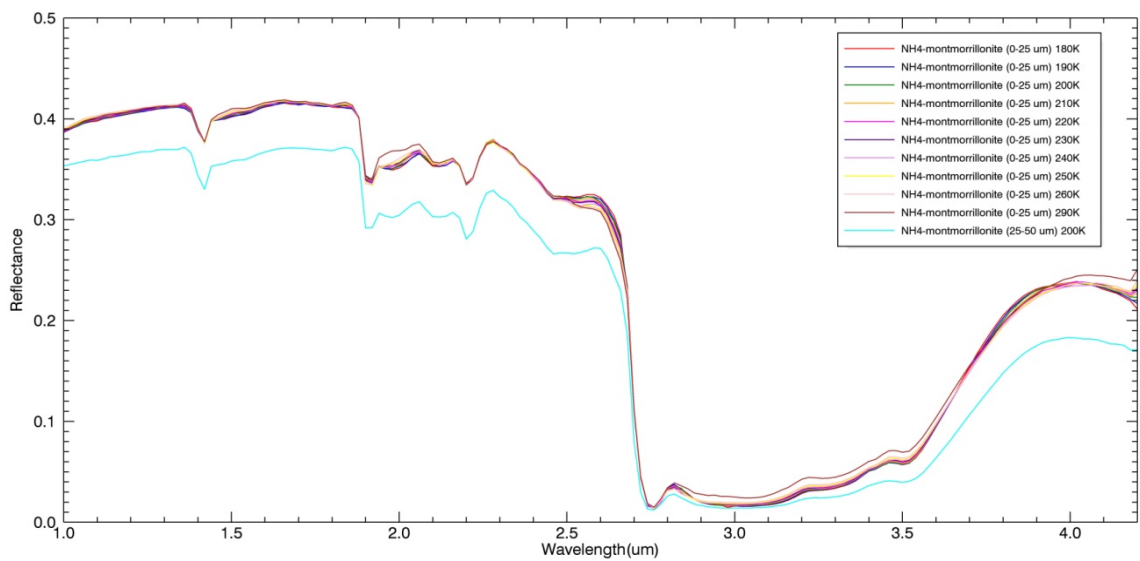

*Figure 2:* Bidirectional reflectance spectra of $NH_4$-montmorillonite (0-25 μm) acquired at different temperatures between 180 and 290 K and the spectrum of $NH_4$-montmorillonite (25-50 μm) acquired at 200 K.

In the **Figure 3** the scatterplot of 2.7 μm band depth vs 3.1 μm band depth is shown for the $NH_4$-montmorillonite sample with grain size 0-25 μm at different temperatures (blue diamonds) and for the $NH_4$-montmorillonite sample with grain size 25-50 μm cooled at temperature of 200 K (red diamond). In *Figure 4*, the scatterplot 3.1 vs 3.4 μm band depth is represented for the same samples. By analysing the spectral features in the NH$_4$-montmorillonite sample with grain size 0-25 μm (blue diamonds) it can be noted that the 2.7 μm band (*Figure 3*) is quite constant at decreasing temperatures (except for an increase in intensity at 210 K and 240 K), whereas the 3.1 μm and the 3.4 μm bands are generally deeper at decreasing temperature (*Figure 4*), as expected *(Freund et al., 1974)* . It is likely that the absorbed atmospheric water influences the 3.1 μm band and at increasing temperature the devolatilization of OH occurs, producing a weaker absorption band.

At fixed temperature (200 K), it is observable that at increasing grain size, i.e. moving (red arrow) from the NH$_4$-montmorillonite sample of 0-25 μm (blue diamond) to the NH$_4$-montmorillonite sample of 25-50 μm (red diamond), the 2.7 μm band becomes deeper and the 3.1 μm band is weaker (*Figure 3*), whereas an increase in the intensity of 3.4 μm band occurs (*Figure 4*). The 2.7 and 3.4 μm bands have larger intensities at coarser size for the high probability of photons to interact with grains and be absorbed, since the volume scattering process dominates in such spectral range *(Salisbury et al., 1993)*. The trend observed for the 3.1-μm band can be explained with a saturation of band, that could be not only due to NH$_4$ but even to absorbed water: the 3.1 μm band depth is weaker at increasing grain size because the absorption in the band center is greater for larger absorption coefficient, so, as the particle size reduces, absorption in the band center remains saturated and a higher reflectance at bands' wings is observed, resulting in an increasing spectral contrast and deeper absorption band *(Myers et al., 2015)*.

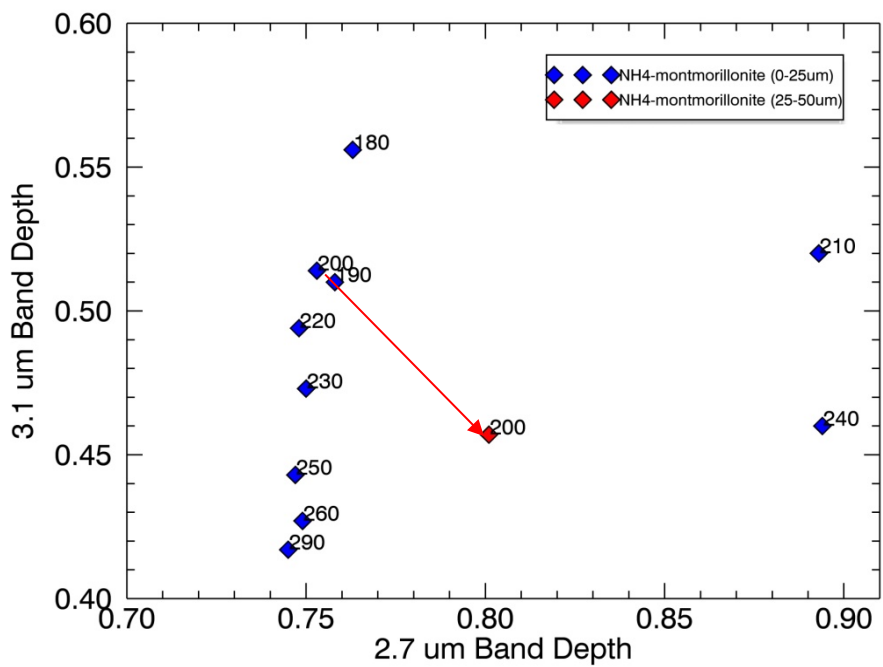

*Figure 3:* Scatterplot of 2.7 vs 3.1 μm BD for the NH$_4$-montmorillonite sample of 0-25 μm grain size cooled at different temperatures (blue diamonds) and for the NH$_4$-montmorillonite sample of 25-50 μm grain size cooled at 200 K (red diamond).

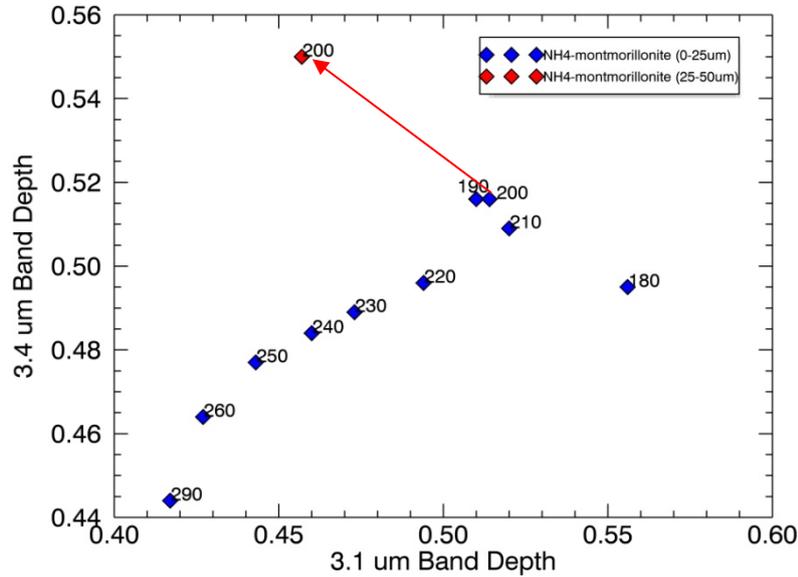

*Figure 4:* Scatterplot of 3.1 vs 3.4 μm BD for the NH$_4$-montmorillonite sample of 0-25 μm grain size cooled at different temperatures (blu diamonds) and for the NH$_4$-montmorillonite sample of 25-50 μm grain size cooled at 200 K (red diamond).

### *4.2 Antigorite*

Antigorite is a hydrous phyllosilicate of Magnesium (Mg) and Iron (Fe), and the spectra of samples analyzed *(Figure 5)* show spectral features related to the OH group, as the 1.4 μm band (overtone O-H stretch $2\nu_3$), the 2.3 μm absorption band (combination tone of Mg-O-H band and O-H stretch) *(Clark, 1999)*, the 2.7 μm band (O-H symmetric stretch $\nu_1$). A spectral feature located at about 2.5 μm is also present, possibly due to Mg-O-H *(Beck et al., 2015)*. The spectra also show absorptions at about 1.9 and 3.1 μm, suggesting a possible absorption of atmospheric water in the antigorite sample. The interest in antigorite sample was limited to the 2.7 μm absorption band, present in the Ceres spectra. The isolation of 2.7 μm band was performed by fitting the spectral continuum with straight lines and the endpoints were estimated in the following ranges: 2.5-2.7 μm and 2.8-2.9 μm.

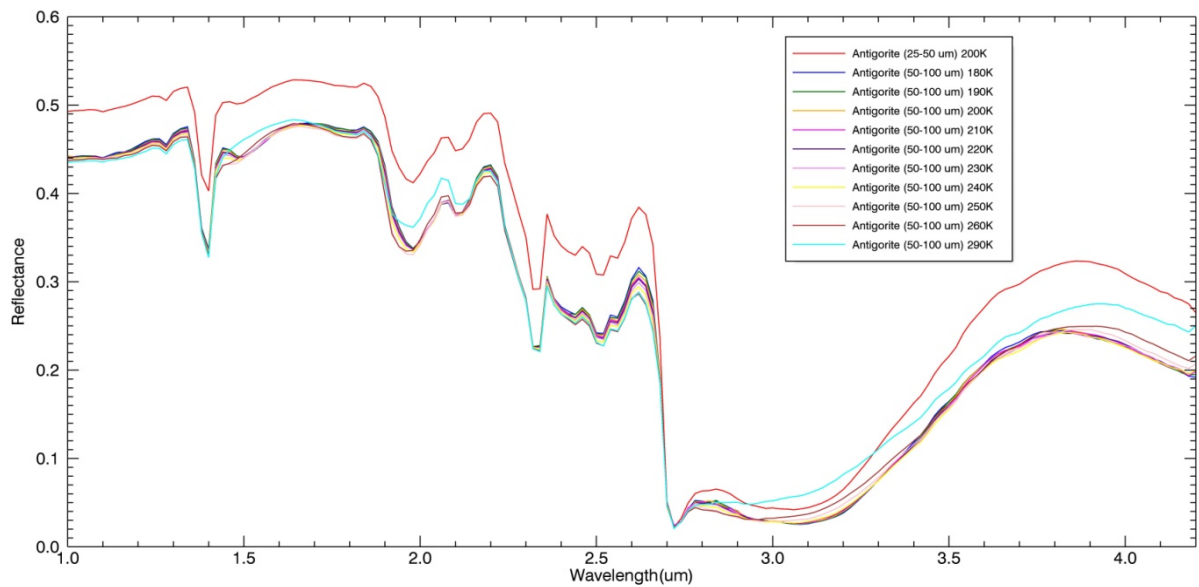

*Figure 5:* Bidirectional reflectance spectra of antigorite sample (25-50 μm) acquired at temperatures included between 180 and 290 K and spectrum of antigorire sample (50-100 μm) acquired at 200 K.

In the *Figure 6*, the scatterplot 2.7 band depth vs temperature is shown for the antigorite sample with grain size 25-50 μm and cooled at 200 K (red diamond) and for the antigorite samples with grain size 50-100 μm cooled at different temperatures between 180 K and 290 K (green diamonds). By taking into account the coarser sample of antigorite and observing the spectral trend of bands' intensity at increasing temperatures (green diamonds), the band depth oscillates at temperatures included between 180 and 220 K. From 220 K to 230 K, the 2.7 μm band decreases and is quite constant to 260 K, whereas the band increases at room T (290 K). No defined trend can be observed between the 2.7 μm band and temperature. By fixing the temperature (200 K), and varying the granulometry, i.e. moving (green arrow) from the 25-50 μm sample (red diamond) to the 50-100 μm sample (green diamond) of antigorite, the 2.7 μm band decreases in intensity, probably related to the saturation of band *(Salisbury et al., 1993)*.

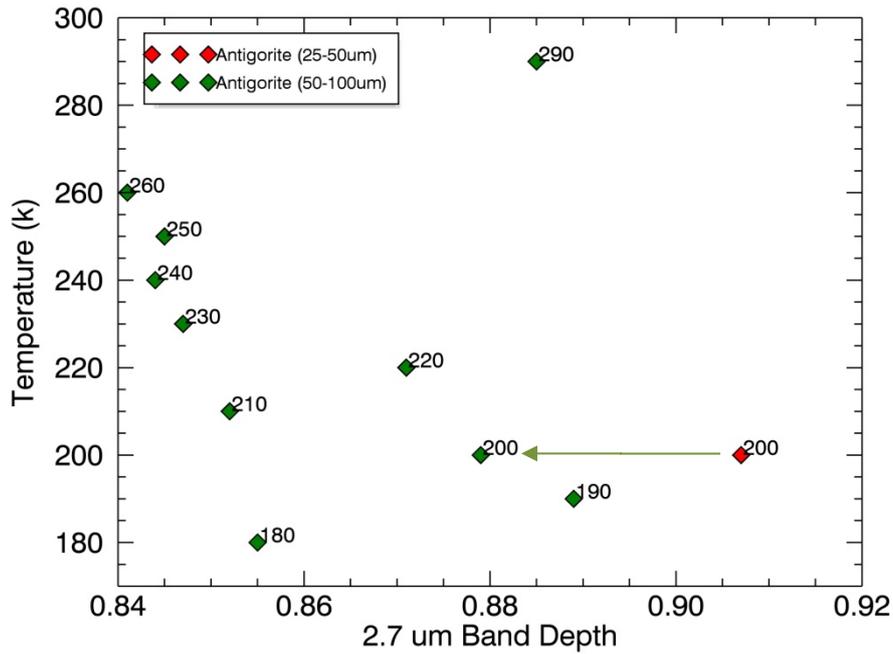

*Figure 6:* Scatterplot of 2.7 band depth vs temperature for the antigorite sample with grain size of 25-50 μm cooled at temperature of 200 K (red diamond) and for the antigorite sample with grain size 50-100 μm, cooled at different temperatures (green diamonds).

*4.3 Dolomite*

The dolomite is a carbonate containing Calcium (Ca) and Magnesium (Mg), the most putative carbonate mineral to compose the Ceres surface *(De Sanctis et al., 2015)*. The fundamental modes of $CO_3^{2-}$ are located at about: 6.4 μm ($CO_3^{2-}$ asymmetric stretch $v_3$), 9.4 μm ($CO_3^{2-}$ asymmetric stretch $v_1$), 11 μm ($CO_3^{2-}$ out-of-plane bend $v_2$), 14 μm ($CO_3^{2-}$ in-plane bend $v_4$) *(Clark, 1999)*. The spectra of the dolomite sample analyzed in this work (*Figure 7*) show absorption bands related to overtone and combination tone of fundamentals *(Clark et al., 1999)*, in particular at about 1.9 μm (due to $2v_1 + 2v_3$), 2.3 μm ($3v_3$), 2.5 μm ($v_1 + 2v_3$), 3.4 μm ($2v_3$) and 4.0 μm ($v_1 + v_3$). Spectral features can be also observed at about 2.7 and 3.1 μm, suggesting a possible contamination due to absorbed atmospheric water. The Ceres reflectance spectra show carbonate absorption bands at about 3.4 μm and 4.0 μm, therefore the analysis of dolomite sample focused on these spectral features. The spectrum of dolomite with grain size 0-25 μm at room T and the spectrum of dolomite with grain size of 50-100 μm at 200 K were acquired.

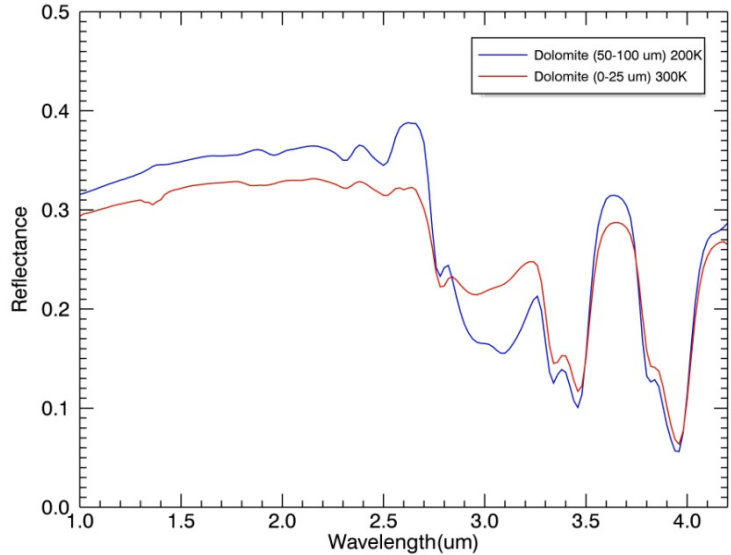

*Figure 7:* Bidirectional reflectance spectrum of dolomite sample (0-25 μm) acquired at room temperature (300 K) and spectrum of dolomite sample (50-100 μm) acquired at 200 K.

The 3.4 and 4.0 μm absorption bands were isolated by fitting the spectral continuum with straight lines and the endpoints were estimated in the following ranges: 3.1-3.2 μm (left wing of 3.4 μm band); 3.5-3.7 (right wing of 3.4 μm band and left wing for 4.0 μm band); 4.0-4.2 μm (right wing of 4.0 μm band). Being very careful in the analysis of dolomite bands, a decreasing in the intensity of both 3.4 and 4.0 μm bands can be observed at increasing temperature and at decreasing grain size (***Figure 8***). The increase of carbonates band's intensity has been previously observed in calcite and dolomite samples at increasing grain size (*Zaini et al., 2012*). The variation in the band's intensity with grain size is related to the fact that coarse grains tend to absorb more radiation penetrating to the grain surface than the fine grains *(Salisbury et al., 1987*; *Clark and Roush, 1984)*, according to Lambert-Beer's Law, typical of transparent material *(Salisbury et al., 1987)*. The decreasing of bands' intensity at increasing temperature in vacuum chamber was also observed in the siderite sample *(Alemanno et al., 2018)*.

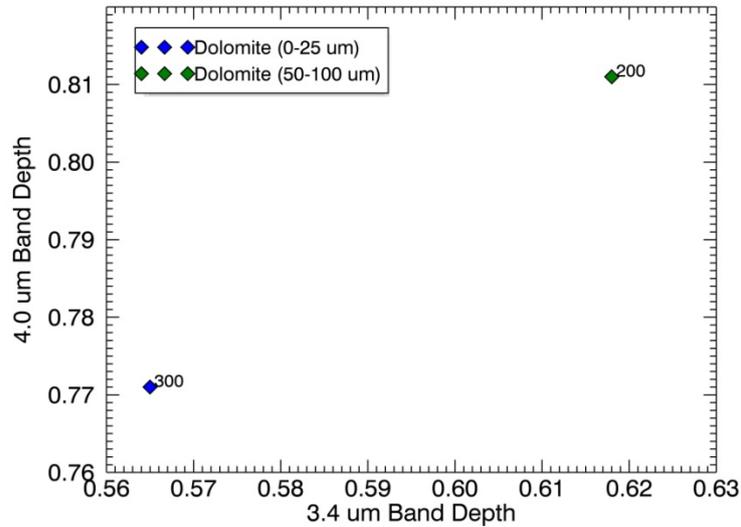

*Figure 8:* 3.4 vs 4.0 μm band depth scatterplot of dolomite sample with grain size of 0-25 μm at 300 K (blue diamond) and dolomite sample of 50-100 μm at 200 K (green diamond).

## 5. Spectral analysis of mixtures

### 5.1 Mixture #1, #2, #3a, #3b, #4, #4_h at room temperature

Based on the Hapke model, the Ceres mean spectrum, as obtained by VIR, is well fitted by a mixture made of 5% Mg-carbonates, 5% antigorite, 6% $NH_4$-montmorillonite and 84% dark component *(De Sanctis et al., 2015)*. The specific carbonate mineral is not fully identified because dolomite, magnesite and calcite produce a similar result, whereas the dark component used in the model is magnetite (*De Sanctis et al., 2015)*.

In this work, we used dolomite as carbonate mineral and graphite as dark component. Graphite is suggested as a darkening agent in addition to magnetite in the composition of Ceres surface: the magnetite as unique dark compound would require an elevated abundance of iron that is not detected on Ceres surface by GRaND *(Prettymann et al., 2019)*. The graphite and the dolomite were mixed with antigorite and $NH_4$-montmorillonite to produce Ceres analogue mixtures: the reflectance spectra of mixtures were first acquired at room temperature (290 K).

Mixture #1 is characterized by end-members grain size of 0-25 μm and the composition reflects that identified by *De Sanctis et al. (2015)*. The graphite is a very dark component and its elevated percentage in the mixture produced, as can be observed in **Figure 9**, a flat and featureless bidirectional reflectance spectrum (blue spectrum), hiding the absorption bands of carbonates and phyllosilicate, that occur in the Ceres mean spectrum (black spectrum). Therefore, we reduced the graphite abundance down to 2% in production of Mixture #2, which has instead an elevated abundance of dolomite (57%) and an abundance of antigorite and the NH4-phyllosilicate of 24% and 17% respectively. The reflectance spectrum of Mixture #2 (violet spectrum in *Figure 9*) is more reflective than the Ceres mean spectrum (black spectrum) and characterized by very deep 2.7, 3.0, 3.4 and 4.0 µm absorption bands. In the Mixture #3a (50-100 µm) we increased the graphite amount (8%), as the dolomite one (64%), and decreased the abundance of antigorite (19%) and $NH_4$-montmorillonite (9%). These variations in the Mixture #3a produced a spectrum (green spectrum in *Figure 9*) with a decrease in reflectance and weaker bands than the Mixture #2, even if the spectrum is still more reflective and the carbonate bands are more intense than the Ceres spectrum. Therefore, just the abundance of graphite (12%) and dolomite (60%) was modified in the Mixture #3a, obtaining the Mixture #3b (grain size 50-100 µm), with a spectrum (yellow spectrum) still too bright with respect to Ceres spectrum and with intense carbonates bands. Finally, the percentage of graphite was increased over again (reaching the 18%), the carbonate percentage was reduced of almost three times (18%) and the abundance of phyllosilicates was increased of about two times for the antigorite (36%) and about three times for the $NH_4$-montmorillonite (28%), obtaining the Mixture #4 (50-100 µm). The reflectance spectrum of Mixture #4 (dark red spectrum) is similar to Ceres spectrum, still characterized by a higher reflectance. Given the evident similarity, a heating process (described in Chapter 2) was performed on Mixture #4, in order to remove the adsorbed atmospheric water, to roughly simulate the solar irradiation experienced by the dwarf planet and obtain a more comparable spectrum to VIR Ceres spectrum. The Mixture #4 after the heating process was termed as Mixture #4_h and it shows a darker reflectance spectrum (light red spectrum) than the Mixture #4.

In the *Table 2*, all the mixtures produced in this work are listed.

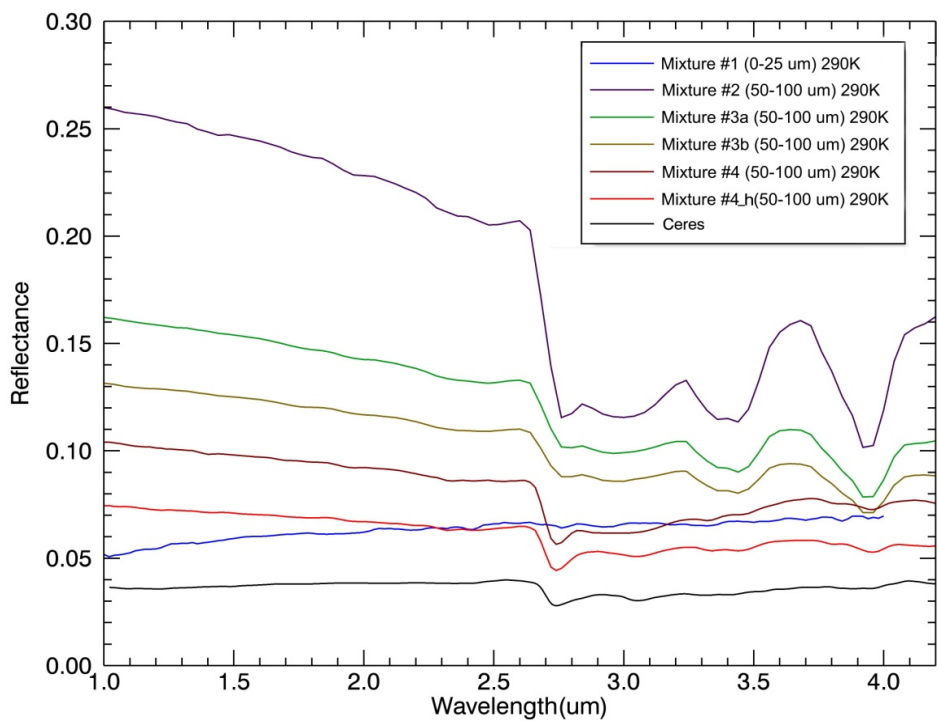

*Figure 9:* Bidirectional reflectance spectra of Mixture #1, Mixture #2, Mixture #3a, Mixture #3b, Mixture #4, Mixture #4_h, compared with the mean Ceres spectrum acquired by VIR.

| Mixture | Grain size (μm) | Dolomite (M%) | Graphite (M%) | Antigorite (M%) | $NH_4$-monrtmorillonite (M%) |
|---|---|---|---|---|---|
| **#1** | 0-25 | 5 | 84 | 5 | 6 |
| **#2** | 50-100 | 57 | 2 | 24 | 17 |
| **#3a** | 50-100 | 64 | 8 | 19 | 9 |
| **#3b** | 50-100 | 60 | 12 | 19 | 9 |
| **#4** | 50-100 | 18 | 18 | 36 | 28 |

*Table 2:* List of Ceres analogue mixtures produced. Each column represents: the name of mixture, the grain size (expressed in μm) and the mass percentage (M%) of each single end-member (dolomite, graphite, antigorite and $NH_4$-montmorillonite).

We compared the obtained mixtures by means of a scatterplot between 2.1 μm reflectance and 1.2-1.9 μm spectral slope (*Figure 10*). Mixture #1 (M1) and the Ceres average show a positive slope, whereas the other mixtures show a negative spectral slope. Anyway, by taking into account the Mixture #2, #3a, #3b and #4 (M2, M3a, M3b, M4 in Figure 10), it can be noted the reflectance decrease at increasing graphite abundance. The heating process performed on the Mixture #4 produced a reflectance decrease and a less negative slope. Such trend was yet observed in laboratory experiments where hydrated carbonaceous chondrites were heated in order to simulate the Ryugu spectrum acquired by NIRS3 onboard Hayabusa2 spacecraft *(Nakamura et al., 2019)*. In particular, the Mixture #4 heated (M4_h) and M1 shows the same reflectance value though the graphite percentage in M1 is about five times than in M4_h. Probably, the heating process produces a partial dehydration of phyllosilicates, with the dark component that is therefore more dominant then in M4 mixture: as result, the reflectance level decreases and the slope approaches the spectral slope of graphite end-member.

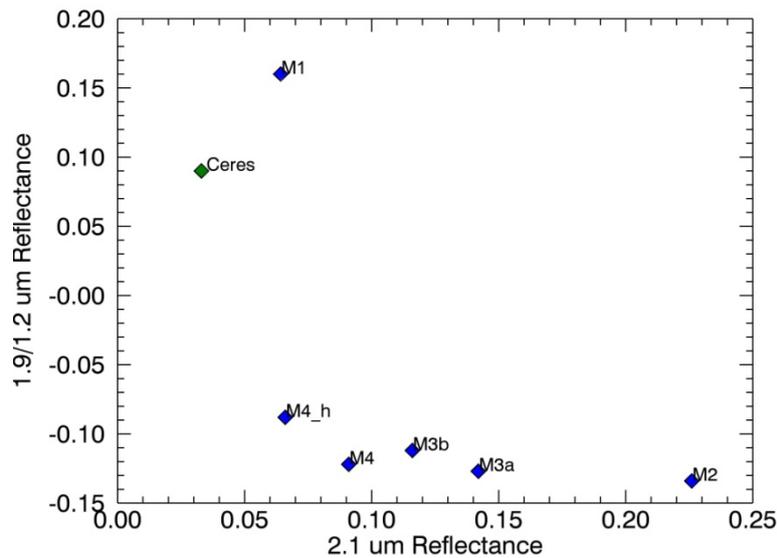

*Figure 10:* Scatterplot of 2.1 µm Reflectance vs 1.9/1.2 µm Reflectance for the Mixture #1, #2, #3a, #3b, #4, #4_h (blue diamonds) and for the mean value of Ceres (green diamond).

In the scatterplot 3.4 vs 4.0 µm band depth (*Figure 11*) the spectral trend of carbonate bands at varying composition is shown. From M2 to M4, the graphite amount increases, producing weaker 3.4 and 4.0 µm bands. In particular, from M2 to M3a the dolomite percentage varies from 57% to 64% but the increase of graphite percentage has a larger influence on the band depths. From M3a to M3b and from M3b to M4 the band shallowing is due to both graphite abundance increasing and carbonate amount decreasing. At fixed end-member's percentage (from M4 to M4h), the heating process produces a slight deepening of 3.4 and 4.0 µm bands, making the M4_h the mixture the most similar to Ceres. The 3.4 µm band is the resulting contribution of both carbonate and $NH_4$-montmorillonite, even if the carbonate probably most effects the 3.4 µm band than the ammoniated mineral. In fact, the $NH_4$-montmorillonite is also responsible of 3.1 µm band, which assumes a different spectral trend with respect to the 3.4 µm band.

The Pearson correlation coefficient (Person) between the 3.1 and 3.4 µm band of mixtures is 0.3, suggesting a weak correlation between these two spectral features. Contrarily, the Pearson coefficient between the 3.4 and 4.0 µm band (carbonate feature) is 0.8, suggesting a strong correlation and therefore, the 3.4 µm band in the mixtures is mainly shaped by dolomite.

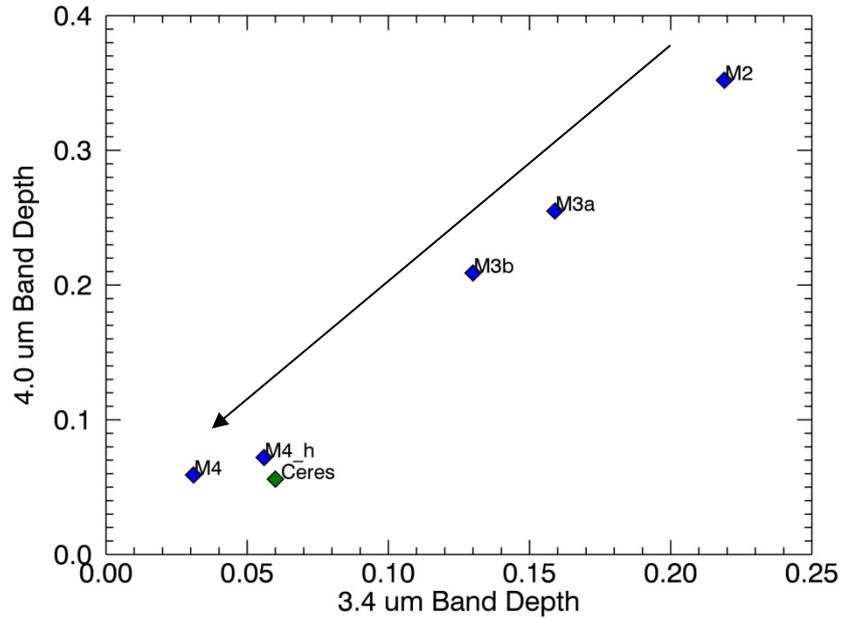

*Figure 11:* Scatterplot of 3.4 vs 4.0 μm Band Depth for Mixture #2, #3a, #3b, #4, #4_h (blue diamonds) and for mean Ceres (green diamond).

In *Figure 12* the 2.7 vs 3.1 μm band depth scatterplot is shown for Ceres analogue mixtures (blue diamonds) and Ceres mean spectrum (green diamond). M3a and M3b are composed by a minor abundance of both antigorite and NH4-montmorillonte than Mixture #2, in addition to a greater percentage of graphite, therefore the 2.7 and 3.1 μm bands are weaker than in M2. Mixture M4 is composed by higher percentage of graphite (18%) than in M3a (8%) and M3b (12%), but the elevated percentage of antigorite (about two times the antigorite percentage in M3a and M3b) and NH4-montmorillonite (about three times the percentage in M3a and M3b) produces deeper 2.7 and 3.1 μm bands than M3a and M3b. Furthermore, the heating process affects the 3.1 and 2.7 μm band, increasing their intensity in the mixture M4_h. The 2.7 μm band is related to OH group of phyllosilicates and the 3.1 μm feature is likely the resulting contribution of both ammonium in phyllosilicates and adsorbed atmospheric water. Therefore, the heating process could have evaporated the absorbed water, or weakened the H-bond interaction between the hydroxyl groups *(Freund et al., 1974)*, so the ammonium remained the only, or the mainly, responsible of the feature. As consequence, the 3.1 μm band in the heated mixture M4_h becomes slightly deeper, assuming the typical shape observed in ammoniated phyllosilicates *(Bishop et al., 2002)* and narrower *(Freund et al., 1974)*, as can be observed in the 3.1 band depth vs 3.1 full width half maximum scatterplot (*Figure 13*). In particular, the 3.1 μm band in the M4_h is almost 2 times narrower than in M4. The more intensity of 2.7 μm band in M4_h is likely a consequence of the loss of absorbed water: the heating process produced a narrower 3.1 μm band and the right shoulder of 2.7 μm feature is no more hidden by the atmospheric water band, so, the spectral continuum of 2.7 and 3.1 um band lies at higher reflectance value, producing an apparent deepening of the 2.7 and 3.1 um bands.

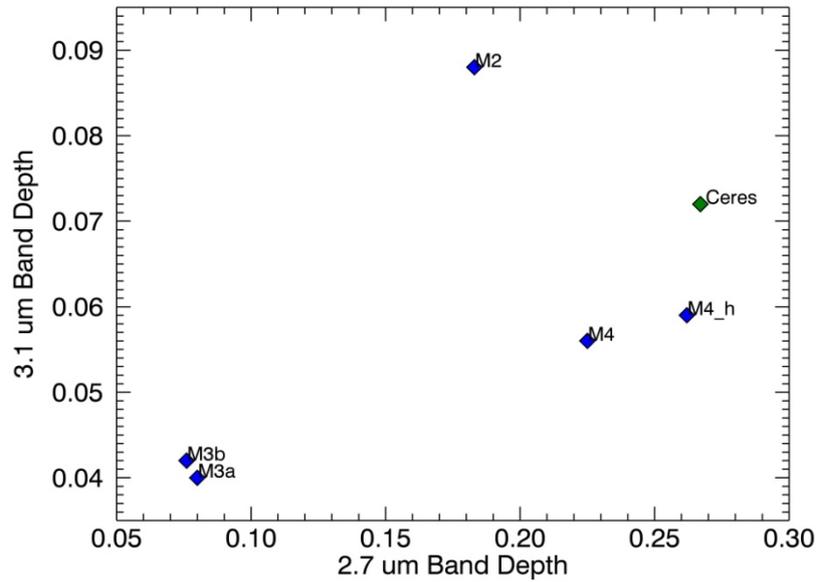

*Figure 12:* Scatterplot of 2.7 vs 3.1 μm Band Depth for Mixture #2, #3a, #3b, #4, #4_h (blue diamonds) and for mean Ceres (green diamond).

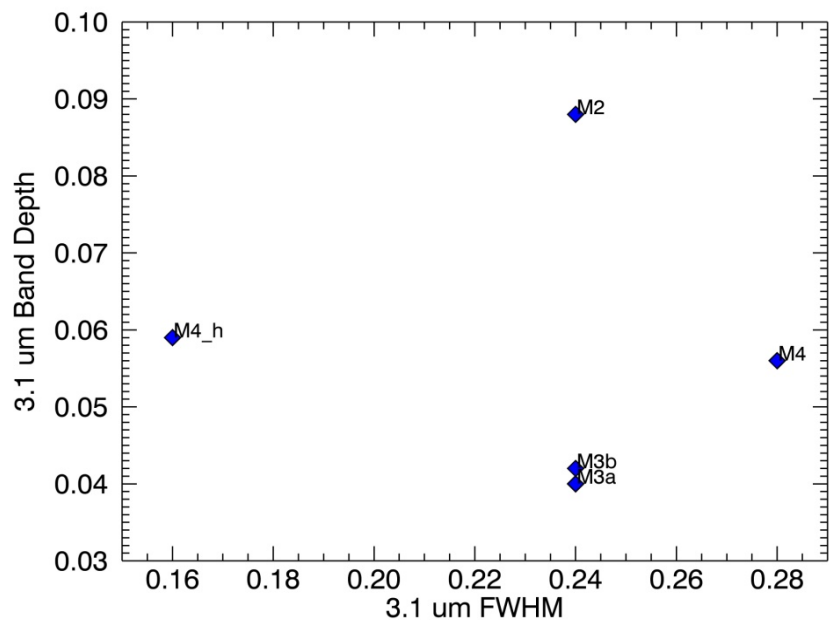

*Figure 13:* Scatterplot of 3.1 μm Band Depth vs 3.1 μm FWHM for Mixture #2, #3a, #3b, #4, #4_h (blue diamonds) and for mean Ceres (green diamond).

Furthermore, the percentage variation of $NH_4$-montmorilloite in the mixtures, as the heating process, influences the band center of 3.1 μm band, as can be observed in *Figure 14*. In M2, M3a and M3b, the 3.1 μm band is likely due to both absorbed water and NH4-montmorillonite but the decreasing percentage of ammoniated mineral in M3a and M3b produces a reduction in the band's intensity and a shift of band center toward shorter wavelengths with respect to M2. Then, the increasing percentage of NH4-montmorillonite in M4 produces a more intense 3.1 μm band and a shift of band center toward longer wavelengths. The heating process, furthermore, removes the absorbed water in the mixture, producing a deeper 3.1 μm band and a band center at longer wavelengths, coincident with the 3.1 μm band center of mean Ceres, i.e 3.06 μm.

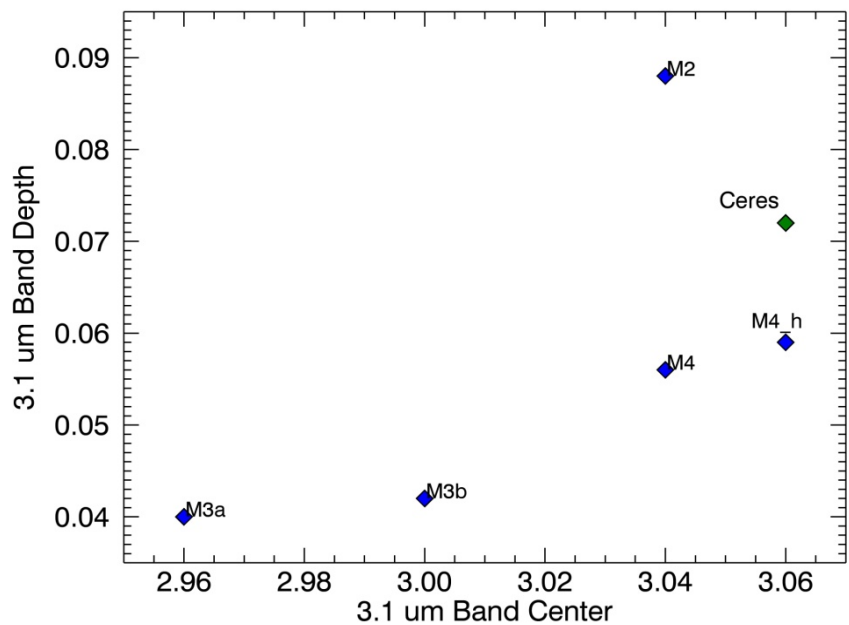

*Figure 14:* Scatterplot of 3.1 μm Band Center vs 3.1 μm Band Depth for Mixture #2, #3a, #3b, #4, #4_h (blue diamonds) and for mean Ceres (green diamond).

The Mixture #4, heated at about 400 K for two hours and composed by grains of 50-100 μm, shows a spectrum, acquired at room temperature (290 K), that is very similar to the mean VIR Ceres spectrum. The spectrum of M4_h (50-100 μm) acquired at 290 K is compared to the mean Ceres spectrum in *Figure 15*. Some spectral differences still occur between the M4_h spectrum (light red spectrum) and the mean Ceres spectrum (black spectrum). Anyway, though the occurrence of some differences in spectral parameters, the Mixture #4 heated is the most similar Ceres analogue mixture among those produced. For this motivation, the same mixture was reproduced with a grain size of 0-25 μm and 25-50 μm and the samples were heated in an oven to remove the adsorbed atmospheric water and to roughly simulate the solar irradiation experienced by Ceres.

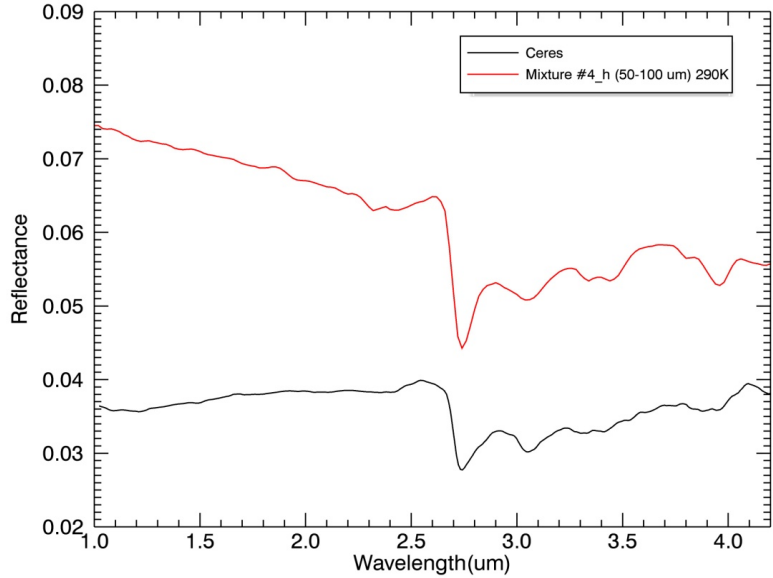

*Figure 15:* Bidirectional reflectance spectrum of Mixture #4_h, with a grain size of 50-100 μm and acquired at 290 K (light red spectrum) compared with the mean Ceres spectrum (black spectrum) obtained by VIR data.

### 5.2  Mixture #4 heated (<25, 25-50 and 50-100 μm) analysed at room and cold temperatures

Since the Mixture #4 heated (Mixture #4_h) shows the most similar spectrum to Ceres, a more detailed spectral analysis of this mixture was performed, by varying the grain size. In particular, reflectance spectra of Mixture #4_h, acquired at room temperature (about 300 K) and with grain size of 0-25, 25-50 and 50-100 μm were compared with spectra of Mixture #4_h, acquired at cold temperature (200 K) and with grain size of 0-25 and 50-100 μm. The spectrum of Mixture #4_h with grain size 25-50 μm was not acquired at cold temperature for unavailability of facility. As can be observed in *Figure 16*, laboratory mixtures are still characterized by bluer slope and by higher reflectance than the mean Ceres spectrum (black spectrum).

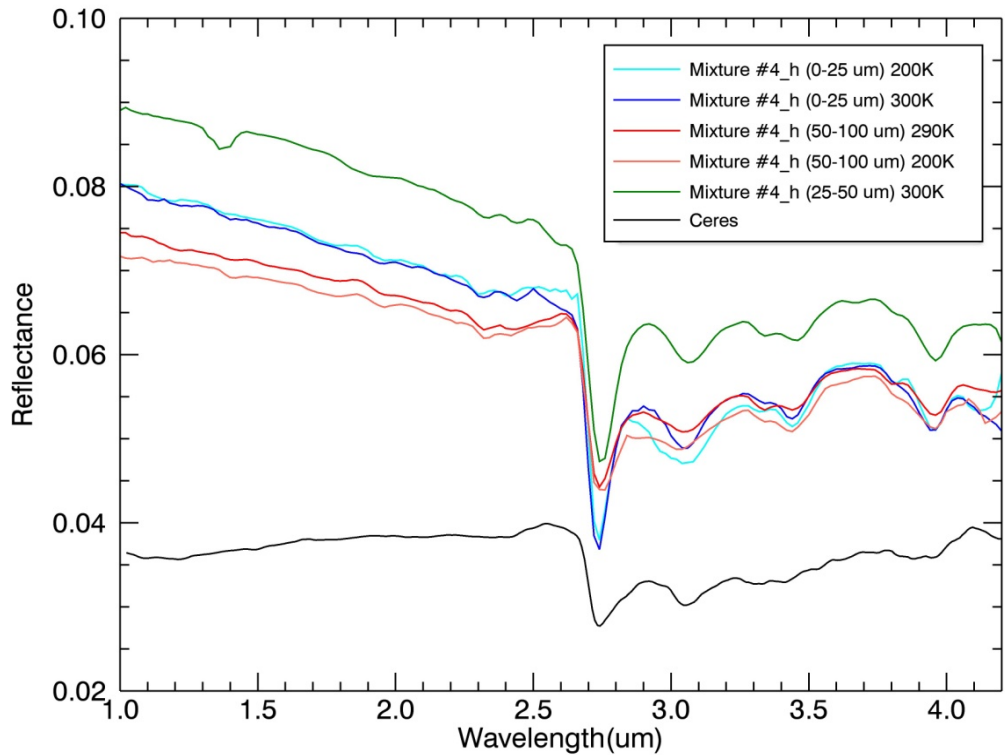

*Figure 16:* Bidirectional reflectance spectra of Mixture #4_h, acquired at room temperature (about 300 K) and with a grain size of 0-25, 25-50 and 50-100 μm, compared with spectra of Mixture #4_h acquired at cold temperature (200 K) with grain size of 0-25 and 50-100 μm. The mean reflectance spectrum of Ceres (black spectrum), obtained by VIR data is also shown for comparison.

In the 2.7 vs 3.1 μm band depth scatterplot (*Figure 17*), it can be noted that the 2.7 μm band reduces in intensity at increasing grain size, both for samples at room temperature (red diamonds) and for samples at cold temperature (blue diamonds), by following the trend observed for antigorite sample (*Figure 6*). Similarly, the 3.1 μm band depth decreases at increasing grain size, both for samples at cold temperature (blue diamonds) and for samples at room temperature (red diamonds), as observed for $NH_4$-montmorillonite sample (*Figure 3*). At fixed grain size, the 3.1 μm band depth does not show a defined trend with temperature and the 2.7 μm band depth slightly decreases at decreasing temperature, assuming an opposed trend to what expected *(Freund et al., 1974)*. By observing the carbonate bands' scatterplot (*Figure 18*), the 3.4 and 4.0 μm bands are weaker at increasing grain size, an opposed behaviour to what observed in the study of dolomite sample (*Figure 8*). Then, a no defined trend in the intensity of carbonate bands is observed at different temperatures when the size of samples is fixed (*Figure 18*).

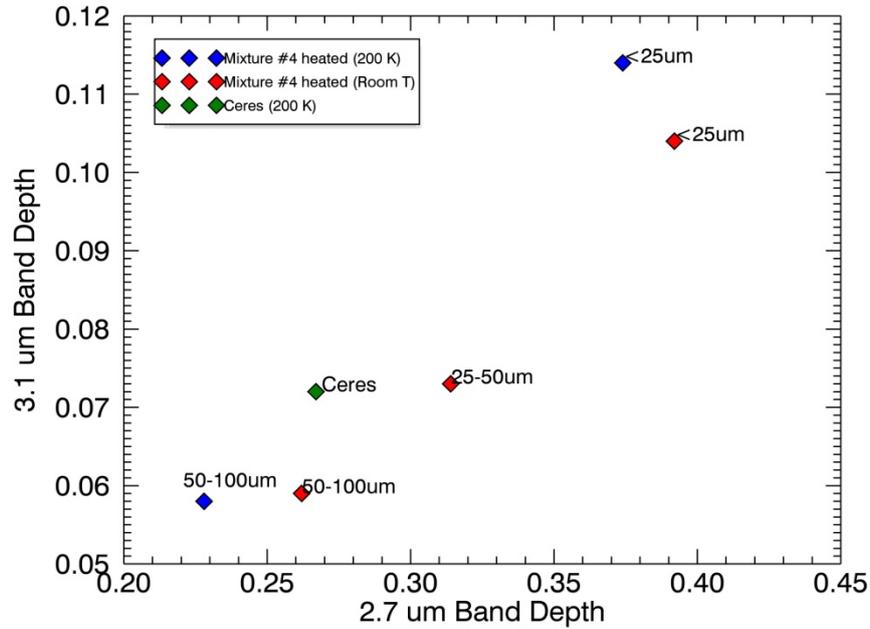

*Figure 17:* 2.7 vs 3.1 μm BD scatterplot for Mixture #4 heated, acquired at 200 K and with grain size of 0-25 and 50-100 μm (blue diamonds) and for Mixture #4 heated, acquired at 300 K and with grain size of 0-25, 25-50 and 50-100 μm (red diamonds), compared with mean Ceres (green diamond).

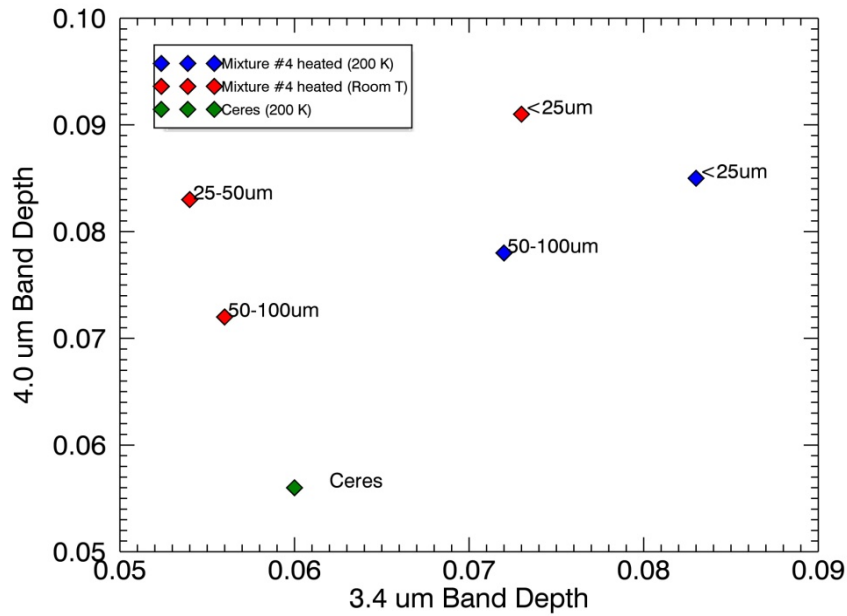

*Figure 18:* 3.4 vs 4.0 μm BD scatterplot for Mixture #4 heated, acquired at 200 K and with grain size of 0-25 and 50-100 μm (blue diamonds) and for Mixture #4 heated, acquired at 300 K and with grain size of 0-25, 25-50 and 50-100 μm (red diamonds), compared with mean Ceres (green diamond).

The slope is also strictly related to the grain size of samples (*Figure 19*). The trend is, however, opposed to what observed in previous laboratory measurements on CI chondrites *(Cloutis et al., 2011)*: Ceres mixtures with grain size of 50-100 μm show a redder slope than Ceres mixtures with finer size, i.e. <25 μm.

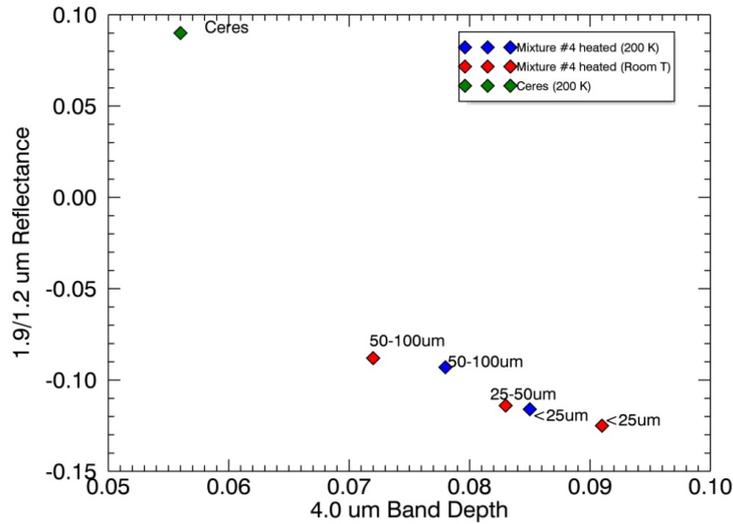

*Figure 19:* 4.0 Band Depth vs slope scatterplot for Mixture #4 heated, acquired at 200 K and with grain size of 0-25 and 50-100 μm (blue diamonds) and for Mixture #4 heated, acquired at 300 K and with grain size of 0-25, 25-50 and 50-100 μm (red diamonds), compared with mean Ceres (green diamond).

Finally, the 3.1 μm band center of Mixtures #4 heated, acquired at room temperature is coincident with Ceres mean spectrum (*Figure 20*), whereas mixtures acquired at cold temperature show a band center at shorter wavelengths: likely, contamination by atmospheric water occurred at low temperatures.

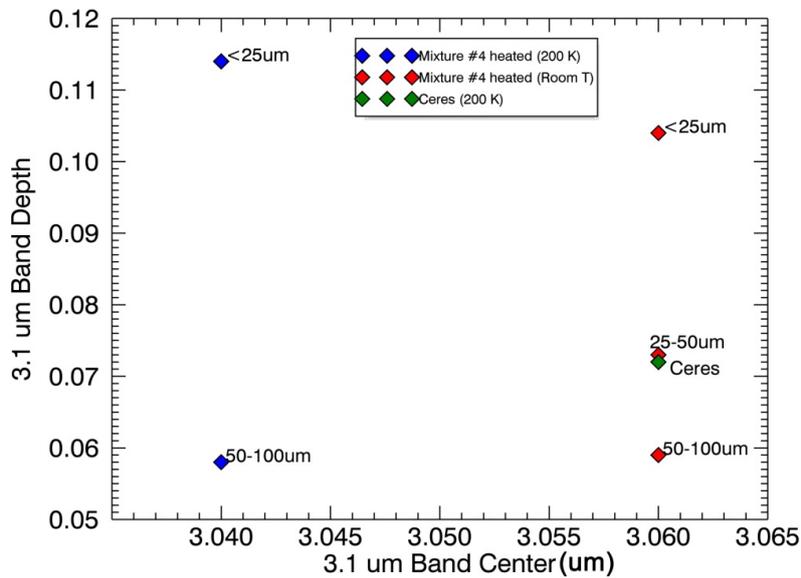

*Figure 20:* 3.1 μm Band Center (expressed in μm) vs 3.1 μm Band Depth scatterplot for Mixture #4 heated, acquired at 200 K and with grain size of 0-25 and 50-100 μm (blue diamonds) and for Mixture #4 heated, acquired at 300 K and with grain size of 0-25, 25-50 and 50-100 μm (red diamonds), compared with mean Ceres (green diamond).

As result of the spectral analysis, the spectra of Mixture #4 heated and acquired at room temperature appear similar to Ceres mean spectrum. In particular, the sample with a grain size of 50-100 μm is spectrally the most similar to Ceres.

## 6. Discussion about the Mixture #4_heated

In the spectral analysis of Mixture #4_h reproduced at different grain size, the 2.7 μm and the 3.1 μm band depths decrease at increasing grain size, both in cold and room temperature, following the spectral trend of the end-members responsible of these spectral features, the antigorite and $NH_4$-montmorillonite sample respectively. Furthermore, the absorptions become weaker at increasing grain size because the size of opaque is also becoming coarser, resulting in a reduction of reflectance spectrum of mixture and in the suppression of absorption features of other minerals *(Salisbury et al., 1987)*. For the same reason, the 3.4 and 4.0 μm bands' intensity decreases at increasing grain size.

The spectral analysis of Ceres analogue mixtures revealed the M4_h as the most Ceres analogue mixture, in particular the Mixture #4 heated with grain size 50-100 μm and with spectrum acquired at room temperature. The spectral parameters of absorption bands are not totally coincident with mean Ceres: the 2.7 and 3.4 μm bands are slightly weaker than mean Ceres; the 4.0 μm band is deeper and the 3.1 μm band is weaker than mean Ceres. Then, the spectrum of M4_h is still brighter than the spectrum of mean Ceres and the spectral slope is negative (opposed trend than in Ceres spectra), though the heating process produced a reddening in the slope.

In the Mixture #4_h obtained at different grain size the spectral slope reveals a more positive value in mixtures composed by coarser grains of end-members. This is a new result, since previous experiments produced mixtures with not coincident grain size of minerals and opaque phase *(Singer, 1980; Cloutis et al., 2011)*. In particular, mixtures of limonite (very fine powder) and magnetite (grain size less than 45 μm), both characterized by a red or flat slope in the 1.2-2.5 μm range, show a blue-sloped spectrum when the end-members are mixed *(Singer, 1980)*. Furthermore, sub-micron magnetite is blue-sloped beyond 0.7 μm *(Morris et al., 1985)* and is red-sloped in coarser size. When the opaque phase as magnetite or graphite is finely dispersed in a mixture, the resulting spectrum is characterized by a blue slope *(Cloutis et al., 2011)*. The graphite show a flat to positive slope in coarse grains (***Figure 1***) and no particular variations are observed in fine size when the graphite is mixed with end-members with similar grain size it is likely that the spectral slope is dominated by the slope of graphite, producing an increasing positive value in coarser samples.

To obtain a more comparable spectrum with Ceres the abundance of dolomite should be decreased and by using dolomite sample with size greater than 100 μm, a weaker 4.0 μm band could be obtained. This could produce a decrease in the intensity of 3.4 μm band, which could be solved by increasing the abundance of $NH_4$-montmorillonite, generating a deeper 3.1 μm band and a better match with the 3.4 μm band. If a sample of 25-50 μm of $NH_4$-montmorillonite is used, no variation in its abundance are necessary. No significant variations occur for the antigorite end-member. The reflectance can be reduced by increasing the abundance of graphite, which could also produce a redder slope: anyway, the reddening of spectrum can be furthermore obtained by using samples with coarser size (in particular dolomite sample with grain size greater than 100 μm) and by applying a heating process to the mixture. The heating process can remove, in addition, the atmospheric absorbed water which alters the 3.1 μm band.

In conclusion, to obtain a more accurate Ceres analogue mixture, it is not sufficient to vary the abundance of end-members, but their grain size should be also taken into account.

Furthermore, a more accurate simulation of space weathering processes experienced by Ceres surface (as micro-meterotic impacts, solar wind irradiation and thermal fatigue) could result in a spectrum with a spectral slope and reflectance similar to Ceres mean spectrum. The Ceres analogue mixtures show a redder spectral slope at increasing grain size, making difficult to explain the negative slope estimated for Haulani ccp as only due to coarse size of end-members. It is anyway plausible that the negative slope in Haulani ccp could be due to the combined effects of reduced space weathering processes and the occurrence of fine grains of dark component in a mixture made of coarse end-members. In particular, laboratory measurements revealed as fine grains of dark component dispersed in a mixture of coarse end-members produce a blue slope in reflectance spectra *(Cloutis et al., 2011)*. Furthermore, the mixtures analyzed in this work were not laser-irradiated or ion-irradiated to respectively simulate micro-meteoritic impacts and solar wind irradiation *(Brunetto, 2009)* on Ceres surface, but just a heating process was performed. Consequently, these mixtures are similar to the "fresher" and less altered material composing the Haulani ccp and their spectra are blue-sloped as the Haulani spectrum *(Figure 21)*. The occurrence of dark component in fine size could produce a bluer spectral slope, approaching the value of Haulani ccp, -0.067 *(Galiano et al., 2019)*.

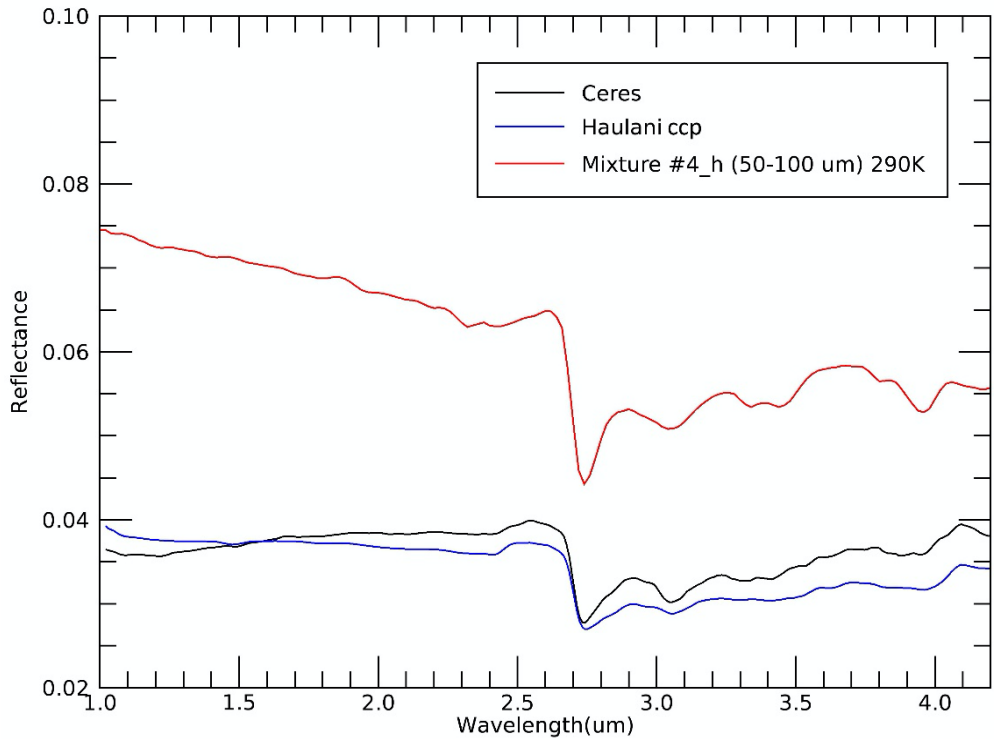

*Figure 21:* VIR spectra of Ceres mean surface (black) and Haulani ccp (blue spectrum) compared with the spectrum of Mixture #4_h (50-100 μm) obtained in this work.

Older complex craters, hence, experienced elevated space weathering processes over time, which fragmented the material composing the peak and mixed the fresher subsurface material with the surface and older one: all these combined effects probably generated a reddening in the spectral slope and a darkening in the spectrum *(Cloutis et al., 2011*; *Palomba et al., 2018; Thangjam et al., 2018).*

## 7. Conclusions

The aim of this work was to study spectral variations of a Ceres analogue mixture with varying temperatures and grain sizes, especially focusing on the latter. Therefore, mixtures composed by different percentage of end-members were produced and spectrally studied, and the Ceres most analogue mixture was detected. In particular, the analysis was performed on the end-members (antigorite, $NH_4$-montmorillonite and dolomite), by studying their spectral behaviour at different grain sizes and different temperatures, and on different Ceres analogue mixtures. After identified the most representative Ceres mixture, it was studied at different grain sizes and temperatures. Although the reproduced mixture is more similar to the Ceres youngest regions (such as Haulani) than to the Ceres average, the endmembers are the same, and the retrieved spectral behaviors can be applied to Ceres study, too.

The results observed from the analysis of end-members are the followings.

NH4-montmorillonite:

- the 2.7 μm band depth does not change at varying temperature, whereas the 3.1 and 3.4 μm bands are weaker at increasing temperatures: probably, the 3.1 μm band is affected by atmospheric water, which devolatilized at relative higher temperatures, whereas the 3.4 μm band follows the trend expected from the literature and the weakening of band is due to the enlargement of interatomic distances at rising temperatures;
- at increasing grain size, the 2.7 and 3.4 μm band depth are deeper and the 3.1 μm band is weaker: for coarser samples of weakly absorbing material the volume scattering dominates, allowing to photons to interact with grains and be absorbed, producing deeper 2.7 and 3.4 μm bands; the spectral behaviour observed for the 3.1 μm band depth can be explained with a saturation of band.

Antigorite:

- the 2.7 μm band intensity does not show a particular trend with temperature;
- at increasing grain size, the 2.7 μm band depth becomes weaker, likely due to saturation of band.

Dolomite:

- the 3.4 and 4.0 μm bands decrease in intensity both at increasing temperature and at decreasing grain size; the spectral trend is anyway uncertain for insufficiency of samples.

The spectral analysis of different Ceres analogue mixtures (Mixture #1, #2, #3a, #3b, #4, #4_h), analyzed at room temperature detected the following results:

- the heating process produces deeper 2.7 and 3.1 μm absorption bands, as consequence of the loss of atmospheric absorbed water; the 3.4 and 4.0 μm also show a deepening in spectral contrast
- the heating process produces a darker spectrum with a redder spectral slope, likely for the devolatilization of OH group in phyllosilicates and a more dominant effect of opaque phase;
- the heating process causes the loss of absorbed atmospheric water in the mixture or the weakening of the H-bond interactions between the hydroxyl group, therefore the $NH_4$-phyllosilicates dominate the 3.1 μm band, making it narrower and with a band center shifted toward longer wavelengths, coincident with the band center of mean Ceres spectrum, 3.06 μm;

- in mixture where both NH$_4$-phyllosilicates and atmospheric water are responsible of the 3.1 µm spectral feature, the band center is shifted toward shorter wavelengths when the percentage abundance of NH$_4$-montmorillonite decreases and is shifted toward longer wavelengths at increasing abundances;
- the mixture #4_h is the most spectrally similar to Ceres spectrum.

The Mixture #4 heated (Mixture #4_h) was reproduced at different grain size and the spectra were acquired at different temperatures, obtaining that:

- the 2.7 µm and the 3.1 µm band depths decrease at increasing grain size, both in cold and room temperature, following the spectral trend of the end-members responsible of these spectral features, the antigorite and NH$_4$-montmorillonite sample respectively; furthermore, the absorptions become weaker at increasing grain size because the size of opaque is also larger, resulting in a reduction in spectral contrast;
- the 3.4 and 4.0 µm bands' intensity decreases at increasing grain size (opposed behaviour to what observed in the spectral trend of dolomite sample, even uncertain) for the large size of opaque grains;
- spectral slope assumes a more positive value in coarser size of grains, likely influenced by the trend of dark component;
- the band center of 3.1 µm band shifts toward longer wavelengths at increasing temperatures: it is likely that spectra acquired at 200 K are affected by atmospheric water, since the band observed is larger as that produced by water's contamination.

In conclusion, the Mixture #4 heated, characterized by a grain size of 50-100 µm and composed of 18% dolomite, 27% graphite, 32% antigorite and 29% NH$_4$-montmorillonite is the mixture more similar to Ceres, among those obtained in this work. In particular, the spectrum acquired at room temperature shows more similarity with the Ceres mean spectrum acquired by VIR data, even if some differences still occur.

On the basis of spectral analysis performed, some suggestions can be provided in order to produce a mixture that is more similar to Ceres spectrum. In particular, variations both in the percentage abundance of end-members and in their grain sizes need to be taken into account.

Furthermore, the space weathering process and the occurrence of dark component in fine size of grains are fundamental to explain the spectral trend observed on Ceres ccps: the negative slope of Haulani, the youngest ccp on Ceres, is probably due to a fresher and weakly processed mixture, composed of fine dark material dispersed among coarse minerals; the redder slope in older ccps is the result of space weathering effects, in particular micro-meteoritic impacts, solar wind irradiations and thermal diurnal variations.

## Acknowledgments

The authors kindly thank Olivier Brissaud for his great contribution in managing experiments and acquiring data.

The measurements described in this work are the outcome of the Trans-National Access research project "VIS-NIR reflectance analysis of Ceres analogue mixture at different grain size to characterize the physical properties of crater central peak material (ccp) on Ceres" selected and funded in the framework of the Europlanet 2020 RI pro- gramme (http://www.europlanet-2020-ri.eu). Europlanet 2020 RI has received funding from the European Union's Horizon 2020 research and innovation programme under grant agreement No 654208.

The whole set of data presented in this paper is available online in the SSHADE database infrastructure (www.sshade.eu). In particular, the doi to obtain spectral data is the following: "Galiano, Anna; Dirri, Fabrizio; Schmitt, Bernard; Beck, Pierre; Brissaud, Olivier (2019): Vis-NIR spectraA. Galiano et al. of NH4-Montmorillonite, Antigorite, Dolomite and Graphite and their mixtures with different grain sizes and temperatures (140-300 K). SSHADE/CSSþREFL_SLAB (OSUG Data Center). Dataset/Spectral Data. doi:10.26302/SSHADE/EXPERIMENT_BS_20191214_001". PB ac- knowledges funding from the European Research Council under the grant SOLARYS ERC-CoG2017-771691.

## References

Alemanno, G. et al., 2018, *Emissivity and reflectance measurements of particulate mixtures for the interpretation of planetary remote sensing data*, European Planetary Science Congress 2018, https://ui.adsabs.harvard.edu/#abs/2018EPSC...12..239A/abstract.

Beck, P. et al., 2010, *Hydrous mineralogy of CM and CI chondrites from infrared spectroscopy and their relationship with low albedo asteroids*, Geochimica et Cosmochimica Acta **74**, pp. 4881–4892, https://doi.org/10.1016/j.gca.2010.05.020.

Beck, P. et al., 2015, *Low-temperature reflectance spectra of brucite and the primitive surface of 1-Ceres*, Icarus **257**, pp.471-476, https://doi.org/10.1016/j.icarus.2015.05.031.

Bishop, J. L. et al., 2002, *Detection of soluble and fixed NH4+ in clay minerals by DTA and IR reflectance spectroscopy: a potential tool for planetary surface exploration*, Planetary and Space Science 50, pp. 11-19, 10.1016/S0032-0633(01)00077-0.

Boer, G.J., Sokolik, I.N., Martin, S.T., 2007, *Infrared optical constants of aqueous sulfate–nitrate–ammonium multi-component tropospheric aerosols from attenuated total reflectance measurements—part I: Results and analysis of spectral absorbing features*, Journal of Quantitative Spectroscopy and Radiative Transfer **108**, pp. 17–38, https://doi.org/10.1016/j.jqsrt.2007.02.017.

Bonnefoy, N., 2001, *Développement d'un spectrophoto-goniomètre pour l'étude de la reflectance bidirectionnelle de surfaces géophysiques. Application au soufre et perspectives pour le satellite Io*, https://cold-spectro.sshade.eu/wp-content/uploads/2015/10/Bonnefoy-Nicolas-2001-these.pdf.

Borden, D. and Giese, R., 2001, *Baseline studies of the Clay Minerals Society source clays: Cation exchange capacity measurements by the ammonia-electrode method*, Clays and Clay Minerals **49**, pp. 444-445.

Brissaud, O. et al., 2004, *Spectrogonio radiometer for the study of the bidirectional reflectance and polarization functions of planetary surfaces. 1. Design and tests*, Applied Optics **43**, Issue 9, pp. 1926-1937, https://doi.org/10.1364/AO.43.001926.

Brunetto, R., 2009, *Space weathering of small solar system bodies*, Earth, Moon and Planets **105**, Issue 2-4, pp. 249-255, https://doi.org/10.1007/s11038-009-9340-9.

Busenberg, E. and Clemency, C. V., 1973, *Determination of the cation exchange capacity of clays ansd soils using an ammonia electrode*, Clays and Clay Minerals **21**, pp.213-217.

Clark, R.N. and Roush, L., 1984, *Reflectance Spectroscopy Quantitative Analysis Technique for Remote Sensing Applications*, Journal of Geophysical Research **89**, pp. 6329-6340, http://dx.doi.org/10.1029/JB089iB07p06329.

Clark, R. N., 1999, *Spectroscopy of Rocks and Minerals, and Principles of Spectroscopy*, pp. 3-52, in Manual of Remote Sensing (Rencz A. N. and Ryerson R. A. Editors).

Cloutis, E. A. et al., 2008, Ultraviolet spectral reflectance properties of common planetary minerals, Icarus 197, pp. 321-347, 10.1016/j.icarus.2008.04.018.

Cloutis, E. A. et al., 2011, *Spectral reflectance properties of carbonaceous chondrites:1. CI chondrites*, Icarus **212**, pp.180-209, doi:10.1016/j.icarus.2010.12.009.

De Sanctis, M. C. et al., 2011, *The VIR Spectrometer*, Space Science Review **163**, pp. 329-369, doi:10.1007/s11214-010-9668-5.

De Sanctis, M. C. et al., 2015, *Ammoniated phyllosilicates with a likely outer Solar System origin on (1) Ceres*, Nature **528**, pp. 241-244, doi:10.1038/nature16172.

De Sanctis, M. C. et al., 2016, *Bright carbonate deposits as evidence of aqueous alteration on (1) Ceres*, Nature **536**, Issue 7614, pp. 54-57, doi:10.1038/nature18290.

Delbo, M. et al., 2014, *Thermal fatigue as the origin of regolith on small asteroids*, Nature **508**, pp. 233-236, doi:10.1038/nature13153.

Ehlmann, B. L. Et al., 2018, *Ambient and cold-temperature infrared spectra and XRD patterns of ammoniated phyllosilicates and carbonaceous chondrites meteorites relevant to Ceres and other Solar System bodies*, Meteoritics and Planetary Science **53**, Nr 9, pp. 1884-1901, doi: 10.1111/maps.13103.

Ermakov, A.I., et al., 2017, *Constraints on Ceres' internal structure and evolution from its shape and gravity measured by the Dawn spacecraft*, Journal of Geophysical Research: Planets **122**, pp. 2267–2293, doi:10.1002/2017JE005302.

Farmer, V., 1968, *Infrared spectroscopy in clay mineral studies*, Clay Minerals **7**, Issue 4, pp. 373-387, doi:10.1180/claymin.1968.007.4.01.

Freund, F., 1974, *Ceramics and thermal transformations of minerals*, Chapter 20, in The Infrared Spectra of Minerals, Farmer, V. C. (editor), pp. 465-482.

Fruwert, J. et al., 1966, *Temperature dependence of the intensity of infrared bands*, Journal of Physical Chemistry **232**, pp. 415-417.

Gaffey, S. J. et al., 1993, *Remote geochemical analysis: elemental and mineralogical composition (Topics in remote sensing)*, Chapter 3, Pieters C.M. and Englert P. A. (editors), pp.43-77, and references therein.

Galiano, A. et al., 2018, *Continuum definition for ∼3.1, ∼3.4 and ∼4.0 µm absorption bands in Ceres spectra and evaluation of effects of smoothing procedure in the retrieved spectral parameters*, Advances in Space Research **62**, pp. 2342-2354.

Galiano, A. et al., 2019, *Spectral analysis of the Cerean geological unit crater central peak material as an indicator of subsurface mineral composition*, Icarus **318**, pp.75-98.

Gillis-Davis, J. J., 2016, *Laser space weathering of possible (1) Ceres analogs*, 47th Lunar and Planetary Science Conference.

Grisolle, F., 2013, *Les condensats saisonniers de Mars: étude expérimentale de la formation et du métamorphisme de glaces de CO2*, Planétologie et astrophysique de la terre [astro-ph.EP], Université de Grenoble, https://tel.archives-ouvertes.fr/tel-01010519.

Harlov, D.E., Andrut, M., Melzer, S., 2001a, *Characterisation of $NH_4$-phlogopite $(NH_4)(Mg_3)[AlSi_3O_{10}](OH)_2$ and $ND_4$-phlogopite $(ND_4)(Mg_3)[AlSi_3O_{10}](OD)_2$ using IR spectroscopy and Rietveld refinement of XRD spectra*, Physics and Chemistry of Minerals **28**, pp. 77–86, http://doi.org/10.1007/s002690000138.

Harlov, D.E., Andrut, M., Pöter, B., 2001b, *Characterization of buddingtonite $(NH_4)[AlSi3O8]$ and $ND_4$-buddingtonite $(ND_4)[AlSi_3O_8]$ using IR spectroscopy and Rietveld refinement of XRD spectra*, Physics and Chemistry of Minerals **28**, pp. 188–198, https://doi.org/10.1007/s00269000014.

Harlov, D.E., Andrut, M., Pöter, B., 2001c, *Characterisation of tobelite $(NH_4) Al_2[AlSi_3O_{10}](OH)_2$ and $ND_4$-tobelite $(ND_4)Al_2[AlSi_3O_{10}](OH)_2$ using IR spectroscopy and Rietveld refinement of XRD data*, Physics and Chemistry **28**, pp. 268–276, https://doi.org/10.1007/s002690000146.

Myers, T. L. et al., 2015, *Quantitative reflectance spectra of solid powders as a function of particle size*, Applied Optics **54**, Nr. 15, pp. 4863-4875, http://dx.doi.org/10.1364/AO.54.004863.

Morris, R.V. et al., 1985, *Spectral and other physicochemical properties of submicron powders of hematite (a-Fe2O3), maghemite (c-Fe2O3), magnetite (Fe3O4), goethite (a- FeOOH), and lepidocrocite (c-FeOOH)*, Journal of Geophysical Research **90**, pp. 3126–3144.

Nakamura, T. et al., 2019, *Possible interpretations of Visible/Near Infrared spectra of asteroid Ryugu obtained by the Hayabusa2 mission*, 50[th] Lunar abnd Planetary Science Conference, #1681.

Neszmelyi, A. and Imre, L., 1968, *Evaluation of IR absorption band intensities in measurements above room temperature*, Spectrochimica Acta **24A**, pp. 297-300.

Parkin, K. M. And Burns, R. G., 1980, *High temperature crystal field spectra of transition metal-bearing minerals: Relevance to remote-sensed spectra of planetary surfaces*, Proceeding in 11th Lunar and Planetary Science Conference, pp. 731-755.

Palomba, E. et al., 2018, *Compositional differences among Bright Spots on the Ceres surface*, Icarus 320, pp. 202-212, https://doi.org/10.1016/j.icarus.2017.09.020.

Pieters, C. M. and Noble, S. K., 2016, *Space weathering on airless bodies*, Journal of Geophysical Research: Planets **121**, pp.1865-1884, doi:10.1002/2016JE005128.

Potin, S. et al., 2018, *SHADOWS: a spectro-gonio radiometer for bidirectional reflectance studies of dark meteorites and terrestrial analogs: design, calibrations, and performances on challenging surfaces*, Applied Optics **57**, Issue 28, pp. 8279-8296, https://doi.org/10.1364/AO.57.008279.

Prettyman, T. H. et al., 2011, *Dawn's Gamma ray and Neutron Detector*, Space Science Review **163**, pp. 371-459, doi:10.1007/s11214-011-9862-0.

Prettyman, T. H. et al., 2019, *Elemental composition and mineralogy of Vesta and Ceres: Distributions and origins of hydrogen-bearing species*, Icarus **318**, pp. 42-55, https://doi.org/10.1016/j.icarus.2018.04.032.

Salisbury, J. W. et al., 1987, *Usefulness of weak bands in midinfrared remote sensing of particulate planetary surfaces*, Journal of Geophyisical Research **92**, pp. 702-701, doi: 10.1029/JB092iB01p00702

Salisbury, J. W., 1993, *Mid-infrared spectroscopy: Laboratory data*, in Remote Geochemical Analysis (Pieters, C. And Englert P. Editors), chap.4, pp. 79-98, Cambridge University Press, New York.

Schmedemann, N. et al., 2016, *Timing of optical maturation of recently exposed material on Ceres*, Geophysical Research Letters **43**, pp. 11987-11993, doi:10.1002/2016GL071143.

Sierks, H. et al., 2011, *The Dawn Framing Camera*, Space Science Review **163**, pp. 263-327, doi:10.1007/s11214-011-9745-4.

Singer, R. B., 1980, The dark materials on Mars: 1. New information from reflectance spectroscopy on the extent and mode of oxidation, Lunar and Planetary Science 11, pp. 1045-1047 (abstract).

Thangjam, T. et al., 2018, *Spectral properties and geology of bright and dark material on dwarf planet Ceres*, Meteoritics and Planetary Science **53**, Nr. 9, pp. 1961-1982, doi:10.1111/maps.13044.

Wendlandt, W. W. And Hecht, H. G., 1966, *Reflectance Spectroscopy*, Wiley Interscience, New York, 298 pp.

Zaini, N. Et al., 2012, *Effect of Grain Size and Mineral Mixing on Carbonate Absorption Features in the SWIR and TIR Wavelength Regions*, Renote Sensing **4**, pp. 987-1003, doi: 10.3390/rs4040987.